\documentclass[twocolumn]{aastex63}

\usepackage{graphicx}
\usepackage{scalefnt}
\received{XXX}
\revised{XXX}
\accepted{December 19, 2019}
\submitjournal{AJ}

\defcitealias{1995ApJS..101...41B}{BS95}
\defcitealias{2008MNRAS.389..678B}{Paper~I}
\defcitealias{2019AJ....157...12B}{BP19}
\defcitealias{2018ApJ...853..104B}{BGB+18}
\defcitealias{1986PASP...98.1113H}{H86}
\defcitealias{2015MNRAS.453.3190B}{BS15}

\shorttitle{Updated SMC and Magellanic Bridge Star Cluster Catalog}
\shortauthors{Bica et al.}

\begin{document}

\title{An Updated SMC and Magellanic Bridge Catalog of Star Clusters, Associations and Related Objects}

\correspondingauthor{Pieter Westera}
\email{pieter.westera@ufabc.edu.br}

\author{Eduardo Bica}
\affiliation{Universidade Federal do Rio Grande do Sul,
Instituto de F\'{\i}sica \\
Av. Bento Gon\c{c}alves 9500,
91501-970, Porto Alegre, Brazil}

\author[0000-0003-3379-994X]{Pieter Westera}
\affiliation{Universidade Federal do ABC,
Centro de Ci\^{e}ncias Naturais e Humanas \\
Avenida dos Estados, 5001,
09210-580, Santo Andr\'{e}, Brazil}

\author[0000-0002-7435-8748]{Leandro de O. Kerber}
\affiliation{Universidade Estadual de Santa Cruz,
Depto. de Ci\^{e}ncias Exatas e Tecnol\'{o}gicas \\
Rodovia Jorge Amado km 16,
45662-900, Ilh\'{e}us, Brazil}
%\affiliation{Universidade de S\~{a}o Paulo,
%Instituto de Astronomia, Geof\'{\i}sica e Ci\^{e}ncias Atmosf\'{e}ricas \\
%Rua do Mat\~{a}o 1226,
%05508-090, S\~{a}o Paulo, Brazil}

\author[0000-0003-4254-7111]{Bruno Dias}
%\affiliation{European Southern Observatory \\
%Alonso de C\'{o}rdova 3107,
%Vitacura, Santiago, Chile}
\affiliation{Universidad Andr\'{e}s Bello, Facultad de Ciencias Exactas, Departamento de F\'{\i}sica, \\
Av. Fernandez Concha 700, Las Condes, Santiago, Chile}
\affiliation{Instituto Milenio de Astrof\'{\i}sica,
Av. Vicu\~{n}a Makenna 4860, Macul, 7820436 Santiago, Chile}

\author[0000-0002-2569-4032]{Francisco Maia}
\affiliation{Universidade Federal do Rio de Janeiro, Instituto de F\'{\i}sica,\\
  Av. Athos da Silveira Ramos, 149, 21941-972, Rio de Janeiro, Brazil}
\affiliation{Universidade de S\~{a}o Paulo,
Instituto de Astronomia, Geof\'{\i}sica e Ci\^{e}ncias Atmosf\'{e}ricas \\
Rua do Mat\~{a}o 1226,
05508-090, S\~{a}o Paulo, Brazil}

\author[0000-0003-1794-6356]{Jo\~{a}o F. C. Santos Jr.}
\affiliation{Universidade Federal de Minas Gerais,
Departamento de F\'{\i}sica, ICEx \\
Av. Antonio Carlos 6627,
31270-901 Belo Horizonte, MG, Brazil}

\author[0000-0001-9264-4417]{Beatriz Barbuy}
\affiliation{Universidade de S\~{a}o Paulo,
Instituto de Astronomia, Geof\'{\i}sica e Ci\^{e}ncias Atmosf\'{e}ricas \\
Rua do Mat\~{a}o 1226,
05508-090, S\~{a}o Paulo, Brazil}

\author[0000-0002-4778-9243]{Raphael A. P. Oliveira}
\affiliation{Universidade de S\~{a}o Paulo,
Instituto de Astronomia, Geof\'{\i}sica e Ci\^{e}ncias Atmosf\'{e}ricas \\
Rua do Mat\~{a}o 1226,
05508-090, S\~{a}o Paulo, Brazil}

\begin{abstract}

We present a catalog of star clusters, associations and related extended objects
in the Small Magellanic Cloud and the Magellanic Bridge
with 2741 entries, a factor 2 more than a previous version from a decade ago. Literature data till December 2018 are included.
The  identification of star clusters was carried out with digital atlases in various bands currently
available in DSS and MAMA imaging surveys.
In particular, we cross-identified recent cluster samples from the VMC, OGLE-IV
and SMASH surveys, confirming new clusters and pointing out equivalencies.
A major contribution of the present catalog consists in
the accurate central positions for
clusters and small associations, including a new sample of
45 clusters or candidates in the SMC and 19 in the Bridge, 
as well as a compilation of the most reliable age and metallicity
values from the literature.
%We also analyzed the recent SMC catalog by Bitsakis and collaborators,
%who detected objects
%with an overdensity search algorithm. They referred to them as clusters,
%but most of them are extended, diffuse, and have low stellar density, 
%so that we classified them as associations.
%We conclude that they found 1175 new objects, whereas 119 have
%counterparts in the literature.
A general catalog must also deal
with the recent discoveries of 27 faint and ultra-faint star clusters and galaxies projected
on the far surroundings of the Clouds, most of them from the DES survey.
The information on these objects has been complemented with photometric,
spectroscopic and kinematical follow-up data from the literature.
The underluminous galaxies around the Magellanic System,
still very few as compared to the predictions from
$\Lambda$ Cold Dark Matter simulations, can bring constraints to galaxy
formation and hierarchical evolution.
Furthermore, we provide diagnostics, when possible,  of the nature of
the ultra-faint clusters, searching for borders of
the Magellanic System extensions into the Milky Way gravitational potential.

\end{abstract}


\section{Introduction}

Star clusters, associations and the field
 stellar population in the Magellanic Clouds
(MC), together with their tidal Magellanic Bridge (MB),
are essential components
to understand the past and future evolutionary stages
of the system as a whole.
The Clouds, together with the Milky Way,
act as a nearby theater of galaxy interactions
\citep{2012MNRAS.422.1957B}.
These different components play a key role in terms of age
distributions \citep{2010AA...517A..50G},
age-metallicity relations \citep{2013ApJ...775...83C},
dynamics \citep{2017MNRAS.467.2980S,2018ApJ...867...19K},
cluster distribution 
\citep[][hereafter \citetalias{2008MNRAS.389..678B}]{2008MNRAS.389..678B},
cluster structure \citep{2014MNRAS.437.2005M}, and
galaxy structure \citep{2001AJ....122..220C}, just to mention a few subjects and studies about them.

The study of the Small Magellanic Cloud (SMC) clusters basically starts with the lists by \citet{1956PASP...68..125K} and \citet{1958MNRAS.118..172L},
with 69 and 116 clusters respectively, where the Kron's objects were included
in the Lindsay's list.
Deeper photographic plates, taken by \citet[hereafter H86]{1986PASP...98.1113H},
provided 213 new relatively faint
clusters, including small associations.
Associations in the SMC were cataloged for instance
by \citet{1985PASP...97..530H},
and \citet[hereafter BS95]{1995ApJS..101...41B}.

Some MB clusters have recently been photometrically studied resulting, as a
rule, in young ages
(\citealp[][hereafter BS15]{2015MNRAS.453.3190B}; \citealp{2015MNRAS.450..552P}).
Associations in the MB are extended with low stellar density 
\citep[][and references therein]{1998AJ....115..154D}.
The field population has also constrained the tidal formation and evolution of the MB
\citep{2017MNRAS.466.4711B,2017MNRAS.471.4571C}, whereas the determination
of the SMC star formation history with the VISTA near-infrared $YJK_s$ survey of
the Magellanic System (VMC) provided an SMC tomography \citep{2015MNRAS.449..639R}.

% BS95 were the first to put together and cross-identify clusters,
\citetalias{1995ApJS..101...41B} were the first to put together and cross-identify clusters,
associations and related objects (hereafter CAROs) in the SMC and MB.
In \citetalias{1995ApJS..101...41B}, 284 new clusters and associations were
also reported. A few years later,
\citet{2000AJ....119.1214B} updated the SMC/MB census.
In \citetalias{2008MNRAS.389..678B}, the SMC and MB were presented together with the LMC CAROs.
\citetalias{2008MNRAS.389..678B} listed 635 star clusters, 385 emissionless associations,
316 associations related to emission nebulae (including supernova remnants, hereafter SNRs),
totaling 1336 entries in the SMC and MB, and 7175 CAROs in the LMC.

The study of CAROs in the Clouds depends on
technological advances, such as high spatial resolution and/or different
spectral domains to probe deeper their contents. Ten years have elapsed since
the last census \citepalias{2008MNRAS.389..678B}, and interesting new
clusters and associations have been identified in this period.
% as a result of
%the observational technologies evolving from digitized photographic plates to CCDs.
Besides, new surveys with
larger telescope apertures and resolving power took place, as well as UV and IR surveys complementing the optical ones
\citep[e. g.][]{2017ApJ...834L..14P,2017AcA....67..363S,2018ApJ...853..104B}.
Finally, the far surroundings of the Clouds were surveyed with
the Dark Energy Survey \citep[DES, e.g.][]{2015ApJ...813..109D}
and complemented with deep follow-up studies \citep[e. g.][]{2018ApJ...852...68C}.
They produced a collection of faint or ultra-faint stellar systems
that challenge our current understanding of the formation and
hierarchical evolution of galaxies
\citep[e. g.][]{2017MNRAS.472.1060D}. On the other hand, these systems are
establishing new landmarks for ultra-faint clusters formed in the Clouds
and kept captive, or dispersed into the Milky Way (MW) potential, as compiled and
discussed in the present paper.
We also point out that nowadays a general catalog of the SMC/MB (and as perspective the LMC)
must include
the stellar clusters that, projected on the celestial sphere, seem extremely far from the MC
barycenter, therefore
constituting an Extended Magellanic System (EMS), in order to better constrain its boundaries.

The new deep photometric survey
VISCACHA\footnote{\url{http://www.astro.iag.usp.br/~viscacha/}}
\citep[VIsible Soar photometry of star Clusters in tApii and Coxi HuguA,][]{2019MNRAS.484.5702M}
is using adaptive optics technology to complement the current and past large surveys
on the Magellanic Clouds.
More specifically, VISCACHA aims at observing the crowded regions of star clusters to get a
complete census of their properties.
 An updated catalog of CAROs in the Magellanic Cloud System will allow
a good target selection and observation efficiency.

The aim of the present study is to
%to gather publications containing CAROs in the last decade,
%and cross-identify them with the previous literature.
collect the published information about the CAROs
in the last decade and to search for new clusters.
One of us (E. B.) inspected Hodge's faint clusters
\citep{1986PASP...98.1113H} and found new similar objects
(SBica in the SMC and BBica in the Bridge) by analysing $J$ (blue) SMC
plates from the UK Schmidt Telescope (Siding Springs, Australia),
scanned with the Machine Automatique \`{a} M\'esurer pour l'As\-tro\-no\-mie (MAMA).
The latter are often referred to as the MAMA/SERC
(Science and Engineering Research Council) plates.
The $BRI$ combined images from the Digitized Sky Survey (DSS) atlas were also analysed.
We end up with an updated general catalog of the SMC and MB clusters.

In Section~\ref{sec:new} we present the observational material
and the cross-identification procedures employed.
We discuss the studies in the present catalog, together with the new discoveries.
In Section~\ref{sec:crossid} we cross-identify objects from previous studies with
the ones from the recent SMC objects catalog by \citet{2018ApJ...853..104B}.
We argue that most of them are associations rather than clusters,
by comparison with the previous literature of associations in the Clouds.
In Section~\ref{sec:explore} we explore the new catalog.
In Section~\ref{sec:AgeMet} we present a compilation of ages and
metallicities of the catalog objects, and analyse them.
In Section~\ref{sec:landmarks} we address the small stellar systems that,
projected on the celestial sphere, seem far from the LMC and SMC,
in view of characterizing an EMS.
Finally, in Section~\ref{sec:conclusion} concluding remarks are given.

\section{New Clusters, Associations and Candidates} \label{sec:new}

The studies on new SMC and Bridge clusters in the last decade are listed
in Table~\ref{tab:studies}, along with three studies prior to \citetalias{2008MNRAS.389..678B}.
%together with three from the previous period.
Column~1 lists the references, Column~2 explains the contents, and Column~3
gives designations or additional information.
These designations are used to list the different object identifications
in our new catalog,
given in Table~\ref{tab:catalog}.
In this Table~\ref{tab:catalog} we provide data not included in \citetalias{2008MNRAS.389..678B},
as well as some corrections:
{\it (i)} SMC SNRs in the MC Chandra
Catalog\footnote{\url{https://hea-www.harvard.edu/ChandraSNR/snrcat_lmc.html}};
{\it (ii)} the acronym GHK \citepalias{2008MNRAS.389..678B} was corrected to GQH \citep{2007ApJ...665..306G};
{\it (iii)} mistakes in \citetalias{2008MNRAS.389..678B} concerning RZ designations \citep{2005AJ....129.2701R}
were corrected.

The following objects from the \citet{1974AJ.....79..858H},
\citet[hereafter B]{1975MNRAS.173..327B} and \citetalias{1995ApJS..101...41B}
catalogs are not CAROs,
and therefore are not included in Table~\ref{tab:catalog}:
{\it (i)} HW7, HW17 and B141 are bright galaxies,
{\it (ii)} H86-65, H86-66, B30 and B84
are galaxies with counterparts in the NASA/NED/IPAC extragalactic database;
and {\it (iii)} BS~1 is a faint galaxy group.
\citetalias{1995ApJS..101...41B} provided a list of faint entries of the B
and \citetalias{1986PASP...98.1113H} catalogs that were doubtful with the
available means at that time. The present analysis using DSS and MAMA $J$
images, retrieved 12 B and 31 \citetalias{1986PASP...98.1113H} clusters or candidates
(Table~\ref{tab:catalog}).

We report some newly discovered faint clusters
and candidates in the SMC (45 objects) and Bridge (19 objects).
The objects were classified from their visual contrast in
the MAMA $J$ images, as illustrated for six of them in
Figure~\ref{mosaic} in the Appendix~\ref{AppNewClusters}.
% We illustrate 6 of them in the Appendix, using MAMA images.}

% Table 1
\begin{deluxetable*}{lll}
\tablecaption{Literature sources used for the cross-identification.\label{tab:studies}}
\tablewidth{0pt}
\tablehead{
\colhead{Reference} & \colhead{Main Contribution(s)} & \colhead{Designations}
}
\decimalcolnumbers
\startdata
\citet{1964MNRAS.127..429W}        & SMC Wing clusters                                            & NGC602-A, NGC602-B \\
\citet{1980IAUS...85..353K}        & association in the Bridge                                    & Kunkel`s Association, KA \\
\citet{2006AA...452..179C}  	   & 3 clusters projected on or related to SNRs                   & CVH \\
\citetalias{2008MNRAS.389..678B}   & departure catalog                                            & Paper~I and references therein \\
\citetalias{2008MNRAS.389..678B}   & tidal dwarf galaxies in the Bridge                           & BS I, BS II, BS III \\
\citet{2009AJ....137.3668C}		   & SMC Wing cluster                                             & NGC602-B2 \\
\citet{2009ApJ...694..367S}		   & small clusters in NGC 346 with HST                           & SGK \\
\citet{2010MNRAS.407.1301B}		   & SMC SNRs, multi-wavelength                                   & SNR \\
\citet{2016MNRAS.460..383P}		   & central SMC IR clusters with VMC in the near-IR              & VMC \\
\citet{2017ApJ...834L..14P}		   & SMC outskirts \& main body with SMASH                        & Piatti or SMASH \\
\citet{2017AcA....67..363S}		   & SMC outskirts \& Bridge with OGLE~IV                         & OGLS, OGLB$^1$ \\
\citet{2018ApJ...853..104B}		   & 1175 new objects (mostly assoc.) in the near-UV and IR & BUS, BIS, BMS \\
Present paper          & 64 new SMC/Bridge clusters with Aladin                 & SBica, BBica \\
Present paper          & updated SMC/Bridge catalog with 2741 entries                 & see present \& previous versions \\
\enddata
\tablecomments{$^1$OGLE clusters have two databases:
(i) the first cluster series was given the acronym SOGLE for SMC clusters \citepalias{2008MNRAS.389..678B}.
(ii) Concerning the recent OGLE-IV cluster series \citep{2017AcA....67..363S},
we employ the acronyms OGLS and OGLB for their SMC and Bridge clusters, respectively,
for the sake of simplicity and space.
Note that SBica 9 and 34 turned out to be duplications of BS13
and BS76, respectively.}
\end{deluxetable*}

\subsection{Cataloging Procedures} \label{subsec:procedures}

The present catalog follows the analysis of its recent MW
counterpart including 10978 CAROs \citep[][hereafter BP19]{2019AJ....157...12B}.
In order to reveal the nature of these objects, we consider:
their positions in equatorial coordinates,
angular sizes, stellar densities, contrast to the field, contaminants, presence
of cluster pairs or multiplets, hierarchical effects, shape and astrophysical
parameters, when available.
Here hierarchy means that one object is included in
another, e.g. a cluster inside an association, so the cluster is
``contained in'' the association.

These procedures were also applied to the \citetalias{1995ApJS..101...41B},
\citet{2000AJ....119.1214B}
and \citetalias{2008MNRAS.389..678B} catalog versions. Compared  with
\citetalias{2008MNRAS.389..678B}, the present data provide deeper material
for the SMC main body and surroundings. In this work,
we employed the DSS $B$, $R$ and $I$ atlases, where
$R$ is the filter most sensitive to atomic line emission, and
$I$ is basically free of emission lines.
The co-added multi-color DSS atlas and the Spitzer co-added bands are deeper.
Particularly deep amongst the newly available surveys are the MAMA/SERC
plates.
%  in the Aladin atlas collection
In the outer parts of the SMC/MB, the recent cluster searches with the
Optical Gravitational Lensing Experiment~IV
\citep[OGLE-IV,][]{2017AcA....67..363S} and
Survey of the MAgellanic Stellar History
\citep[SMASH,][]{2017ApJ...834L..14P}
are in general deeper than the DSS (available via the
Aladin\footnote{\url{https://aladin.u-strasbg.fr/}} software).
In this case we cross-identified and incorporated them.

Table~\ref{tab:catalog} includes 1447 entries corresponding to the updated
literature, including the ones from the \citet{2018ApJ...853..104B}
catalog, which are treated in Section~\ref{sec:crossid}.
Column~1 provides the designations in chronological order,
so that discoveries can be verified.
Re-discoveries are not a demerit, since they reinforce an object detection
independently by different authors \citepalias{2019AJ....157...12B}.
Columns~2 and 3 give the J2000 right ascension (R. A.) and declination (Dec.),
respectively.
Compared to \citetalias{2008MNRAS.389..678B}, we now provide the time second decimal of the R. A.
We measured this value for essentially all clusters and small associations.
Earlier SMC and LMC catalogs  were based on photographic plates obtained
by different authors who derived approximate coordinates.
The Digitized Sky Survey plates with astrometry started to change that to
a new paradigm
\citep[][and references therein]{2008MNRAS.389..678B}.
Nowadays, Aladin makes available digital surveys, either from
plates, CCD or other detector surveys with good astrometric accuracy.
However, crowding and saturation effects inhibit attempts to find centers
automatically by stellar statistical techniques or flux peak fits, such that
in some recent studies based on automatic searches, the
coordinates may correspond to off-center positions.
For detailed barycenter studies, higher resolution observations are needed,
e.g. with SOAR/SAM from the ground, or with HST.
Visual inspection on survey images is a reliable method to systematically
estimate cluster centers for catalogs,
in particular in cases of crowded fields.
In the present analysis, all the clusters have centered coordinates.
Since for large associations and stellar/nebular complexes this time second decimal
becomes irrelevant, we simply appended zero as decimal to such
\citetalias{2008MNRAS.389..678B} objects.

The object classes in column~4 (C, A, CA, AC, NA, AN, NC, CN, EN)
and SNR are the same as defined in \citetalias{2008MNRAS.389..678B},
and are explained in Table~\ref{tab:census}.
For more details on this classification, see \citetalias{2008MNRAS.389..678B}.
A new class is added: ``CC'' meaning ``cluster candidate''.
The catalog also contains three tidal dwarf galaxies or ``TDG'' \citepalias{1995ApJS..101...41B}.
The number counts of these objects in the present catalog are also given
in Table~\ref{tab:census}.

Major and minor angular sizes in Columns~5 and 6 are guiding values measured
by ourselves, estimated visually directly on the plates, or directly
taken from other studies with deeper observations,
which in general follow similar procedures to measure diameters.
The objective is to provide basic information to compare the objects
in view of selection criteria for future detailed studies.% \edit1{for example}.
Column~7 refers to the present classifications of the
\citet{2018ApJ...853..104B} objects as defined in Section~\ref{sec:crossid}.
Columns~8 and 9 give the ages and metallicities compiled as
described in Section~\ref{sec:AgeMet}, and Columns~10 and 11 list
the corresponding references.
Comments in Column~12 provide additional information
such as hierarchical relations (e.g. ``in'' or ``include'')
or whether the object appears in a pair or multiplet, as for example 
a cluster present in an association).

During the verifications of new literature objects
in DSS and MAMA $J$ images,
one of us (E. B.) checked Hodge's faint clusters \citep{1986PASP...98.1113H}.
During this verification,
 new similar clusters and cluster candidates were detected,
using MAMA and the color $BRI$ combined images in Aladin:
45 in the SMC, which we named SBica, and 19 in the Bridge area,
analogously named BBica.
These discoveries are incorporated in Table~\ref{tab:catalog}
and some of them are shown in Figure~\ref{mosaic} in Appendix~\ref{AppNewClusters}.
%, using the new acronyms ``S'' and ``B'' representing clusters
%in the SMC and Bridge regions, respectively.

\citet{2012MNRAS.425.3085P} analyzed frames from the Blanco 4~m telescope,
obtained with a CCD camera equipped with Washington
filters to study \citet{1986PASP...98.1113H} faint cluster candidates
in the SMC central bar.
Part of them were confirmed not to be clusters
by means of color-magnitude diagrams (CMDs). We indicate them as ``Ast'' in the comment field
(Table~\ref{tab:catalog}), indicating their probable nature as asterism.
However, it would be important to observe them deeper because
they may be counterparts of Galactic open clusters,
not yet sampled in large numbers in the Clouds.
We recall that \citet{1998MNRAS.295..860S} detected two faint counterparts
of MW open clusters using serendipitous HST observations of a rich field
on the east side of the LMC bar.

\section{Cross-identification with the Bitsakis et al. SMC Catalog}\label{sec:crossid}

\citet[hereafter BGB+18]{2018ApJ...853..104B}, provided the largest sample
of SMC objects in the last decade (Table~\ref{tab:studies}).
We cross-identified their objects with the literature (Section~\ref{sec:new}).
They employed a code that automatically detects overdensities above a local
threshold.
Monte-Carlo simulations probed the background and the code detected both
compact and diffuse overdensities.
They calculated their ages by CMD fitting in the $(U-V)$ vs. $V$, $(B-V)$ vs. $V$,
and $(V-i)$ vs. $i$ diagrams.
However, for older clusters the data they use do not reach the turn-off,
resulting in uncertain age determinations.
They analysed the following three databases: {\it (i)} SMC main body with GALEX in
the near-UV ($\lambda_{\rm eff}=2175$~\AA);
{\it (ii)} central parts of the SMC in the Swift/UVOT Magellanic Clouds Survey
with the near-UV filters UV $W1$, $W2$ and $W3$; and
{\it (iii)} the SMC main body with Spitzer/IRAC $3.6~\mu{\rm m}$.
They designated the objects with the acronyms SMC-NUV, SMC-M2 and SMC-IR1,
respectively.
For the sake of simplicity and space, we abbreviated them
in the present catalog to BUS, BMS and BIS, respectively.
The ``B'' in these acronyms refers to Bitsakis and ``S'' to the SMC,
as usual in several catalogs (Table~\ref{tab:studies}, \citetalias{2008MNRAS.389..678B}).

\citetalias{2018ApJ...853..104B} referred to their detected objects as star clusters.
The publication of such a cluster sample in excess of 1000 entries was
surprising, and it would have an enormous impact on cluster luminosity functions.
\citet{2018MNRAS.478..784P} argued that the unprecedented number
of new clusters could be greatly overestimated.
In order to clarify this issue, we inspected the \citetalias{2018ApJ...853..104B}
objects taking into
account the procedures in Section~\ref{subsec:procedures}, and determining
their angular separations to known objects from the literature and to each other.
We searched for counterparts of the objects in \citetalias{2018ApJ...853..104B} to test the
reliability of this selection and to collect additional information for the
nature of these objects. The counterparts were verified using a number of
criteria, including angular separation,
diameters, classifications, and checking the DSS and MAMA $J$ images.
Most of them are located at less than 60~arcsec from known objects.
Their decreasing number for separations larger than that ensures
that we tested the bulk of near coincidences in positions.
The Bitsakis objects were here classified into:

\begin{itemize}
\item {\it (i)} 961 type ``I'' corresponding to isolated objects in Column~7 of
Table~\ref{tab:catalog}. As a rule they are extended, diffuse and with low stellar
densities, corresponding to properties of
associations in the Clouds \citep[e. g.][BS95]{1985PASP...97..530H}.
In particular, they do not correspond to a typical faint cluster appearance
\citep{1986PASP...98.1113H}.
We conclude that such objects are to be classified as associations.
In fact, many of them are not clearly seen on DSS or MAMA images,
such that we cannot exclude the possibility that they are field fluctuations.
This might be due to the fact that they used particular near-UV
and IR material, having detected overdensities therein, but
have no clear counterpart in the optical.
Assuming these objects to be real, we decided to include
all such \citetalias{2018ApJ...853..104B} objects in the association class, which are readily
discernible in our Table~\ref{tab:catalog}, column~4;
\item {\it (ii)} 214 objects build pairs with other objects from our catalog,
but are not similar enough to these to be considered
the same object.
% !!! O referee quer mais detalhes sobre "not similar"
We classify them as ``N'', referring to non-similar;
\item {\it (iii)} 119 objects have counterparts in the literature with
comparable size and description.
For this class we use the designation ``E'', which stands for equivalent.
Most of them are previously cataloged bright and moderately bright
compact clusters. For the first time, the names of \citetalias{2018ApJ...853..104B}
objects with counterparts in the literature are explicitly given
in the same catalog line, as suggested by \citet{2018MNRAS.478..784P}. Finally,
we detected some equivalencies among objects from their three databases
(BUS, BMS and BIS), and to a lesser extent within the same database.
These internal duplications are included in Table~\ref{tab:catalog};
\item {\it (iv)} 1351 objects in our catalog have no relation to any object
in the \citetalias{2018ApJ...853..104B} sample. We classify them as ``U'' (unrelated);
\item {\it (v)} yet, 96 objects are hierarchically related to the
\citetalias{2018ApJ...853..104B} objects, so we classified them as
``R'', meaning related entries.
\end{itemize}

In conclusion, the diffuse \citetalias{2018ApJ...853..104B} objects amount to 1175
(43\% of the present overall catalog), corresponding mostly to associations.
Their 119 compact objects are previously cataloged bright and moderately 
bright clusters. In essence, they have no faint clusters.
Their determinations indicate a considerable
fraction of ages over 100~Myr, thus older than typical OB associations.
This suggests the occurrence of evolved associations and/or cluster dissolutions. 
Finally, the entries BIS767 and BUS486 were excluded because they
are part of the Milky Way globular cluster NGC\,362.

We point out that the objects in Table~\ref{tab:catalog} span a wide
range in size, classes and stellar or gas content.
The previous literature cited from \citetalias{1995ApJS..101...41B}
to the present paper shows definitions and images of the different classes.
We suggest the use of %Aladin bands with
the present accurate coordinates
and other characterizations for preliminary analyses to select object samples
for observations.

\section{Properties of objects in the updated Catalog} \label{sec:explore}

% Fig. 1
\begin{figure*}[ht!]
%\plotone{SMCcatFig1.eps}
%\includegraphics[width=1.0\textwidth]{SMCcatFig1.pdf}
%\includegraphics[width=1.0\textwidth]{SMCcatFig1.eps}
\includegraphics[width=1.0\textwidth]{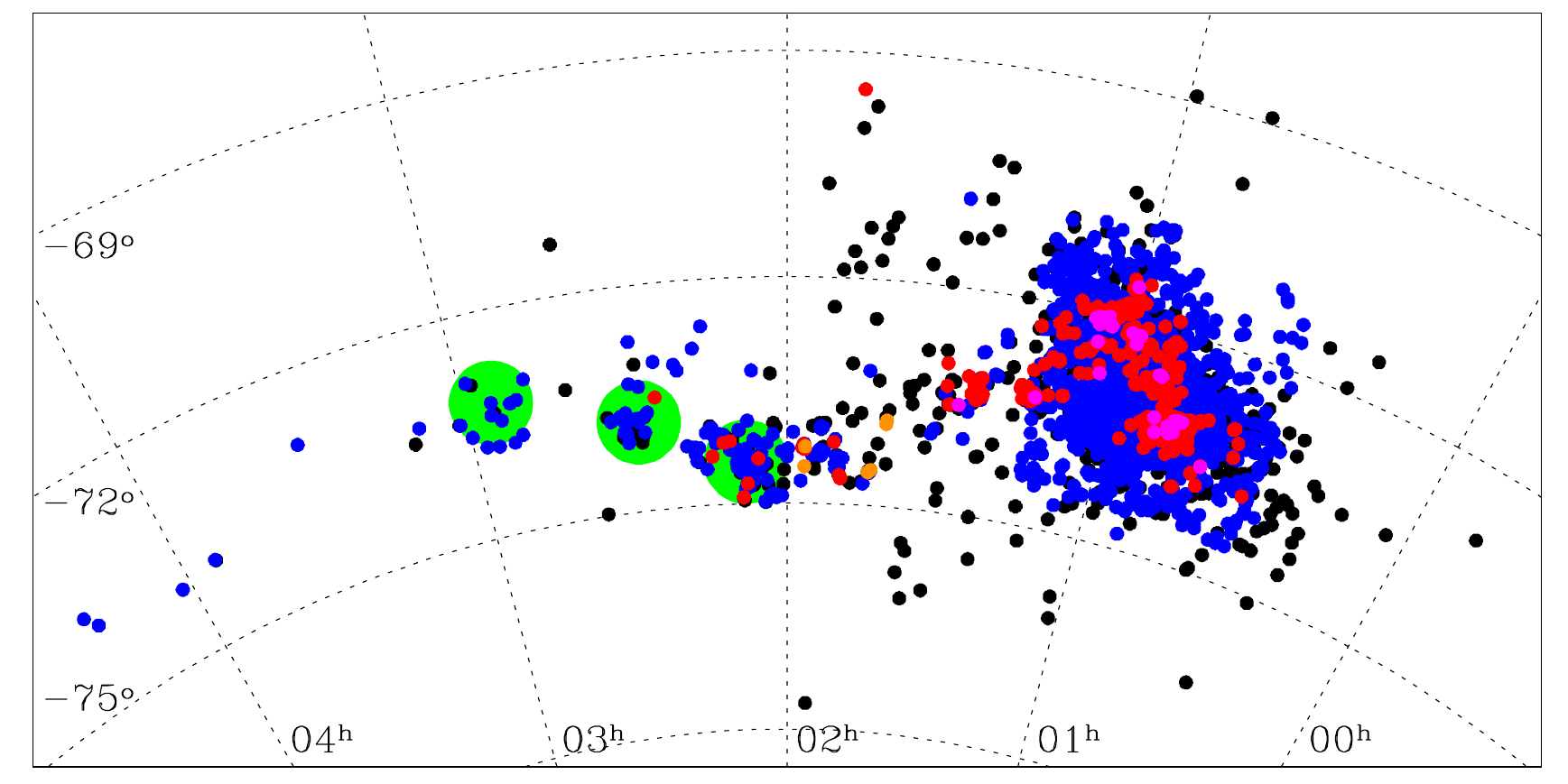}
  \caption{Angular positions of the general catalog objects. Points are
(i) black: star clusters (C, CA, CC),
(ii) blue: associations without emission (A, AC),
(iii) red: clusters and associations related to emission (NC, NA, AN, CN),
(iv) magenta: SNRs, (v) orange: ENs,
and (vi) finally, the three TDGs as large green circles.\label{fig:objtypes}}
\end{figure*}

\addtocounter{figure}{-1}

\begin{figure}[ht!]
%\plotone{SMCcatFig1b.pdf}
%\plotone{SMCcatFig1b.eps}
\plotone{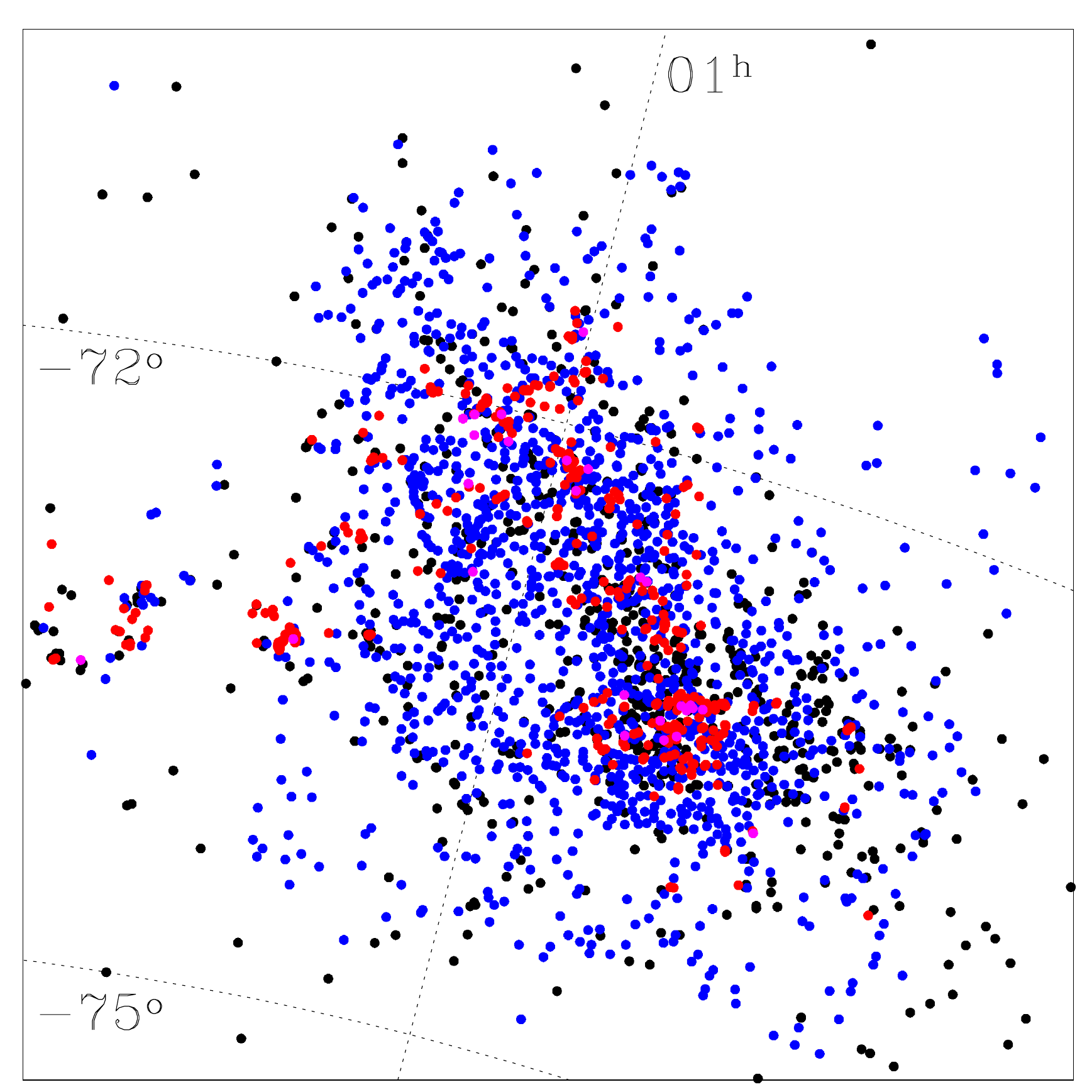}
\caption{{\it (continuation)} Enlargement of the most crowded region of
Figure~\ref{fig:objtypes}, the SMC main body, with smaller points size.}
\end{figure}

% Fig. 2
\begin{figure*}[ht!]
%\plotone{SMCcatFig2.eps}
%\includegraphics[width=1.0\textwidth]{SMCcatFig2.eps}
%\includegraphics[width=1.0\textwidth]{SMCcatFig2.pdf}
\includegraphics[width=1.0\textwidth]{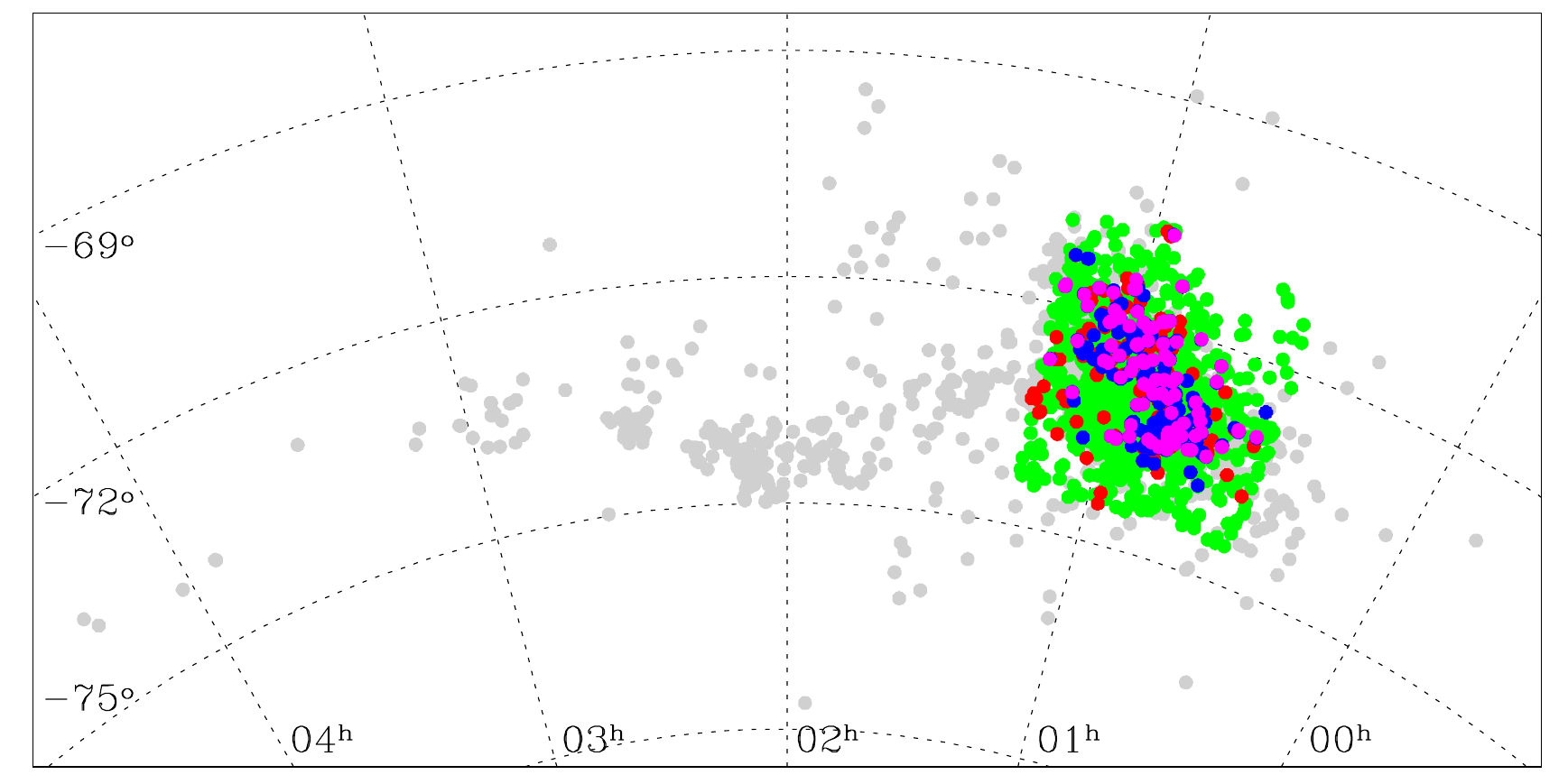}
\caption{Comparison with \citetalias{2018ApJ...853..104B}:
(i) grey points: present sample,
(ii) green: 961 isolated objects (new associations),
our \citetalias{2018ApJ...853..104B} catalog correlation class ``I'',
(iii) blue: their 214 objects, mostly associations,
separated from literature entries by less than 60 arcsec, class ``N'',
(iv) red: their 119 clusters with counterparts in the literature,
class ``E'',
and (v) in magenta: 96 SMC objects in the literature not equivalent,
but apparently related to \citetalias{2018ApJ...853..104B} objects, class ``R''.\label{fig:corclasses}}
\end{figure*}

\addtocounter{figure}{-1}

\begin{figure}[ht!]
%\plotone{SMCcatFig2b.eps}
%\plotone{SMCcatFig2b.pdf}
\plotone{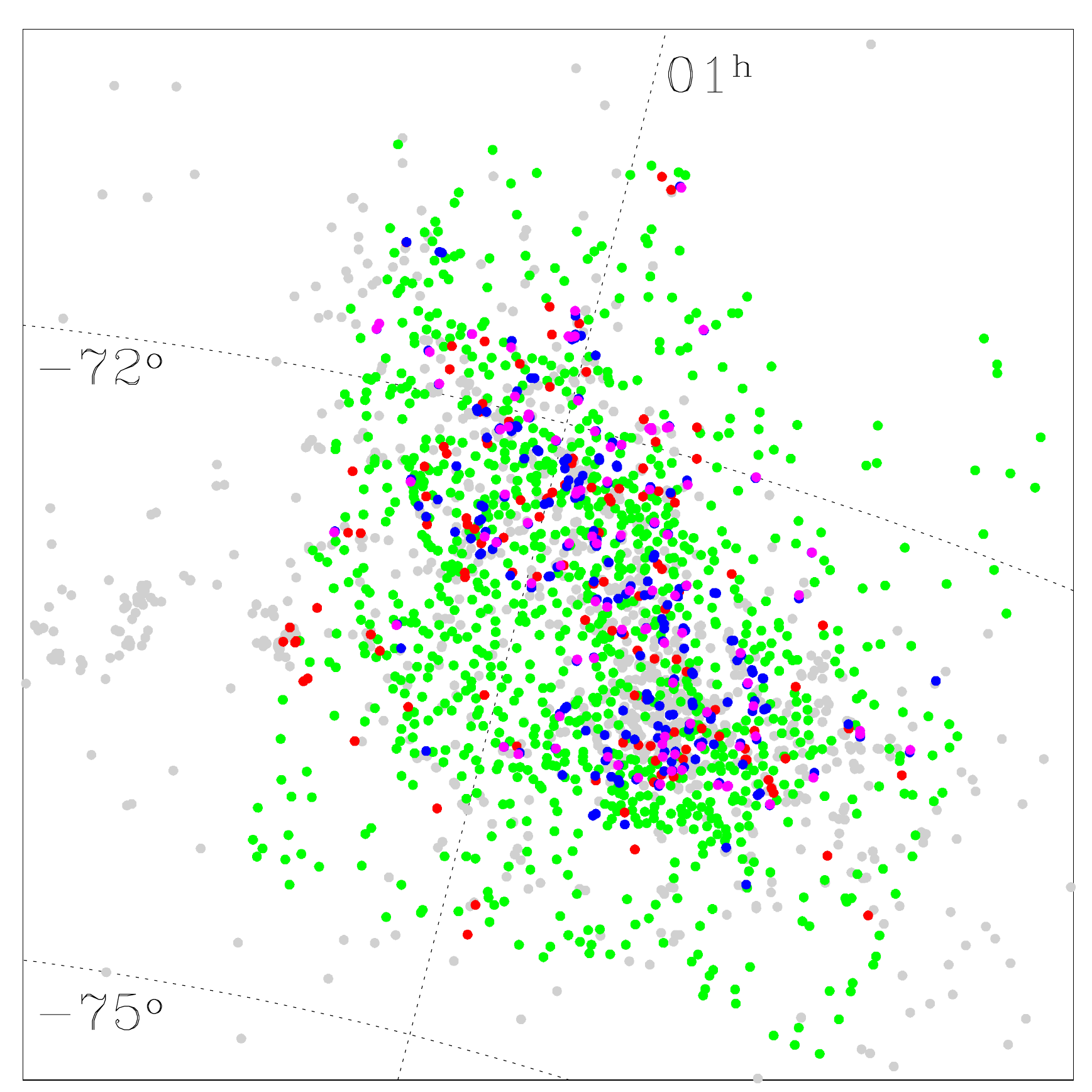}
\caption{{\it (continuation)} Enlargement of the most crowded region of
Figure~\ref{fig:corclasses}, the SMC main body, with smaller points size.}
\end{figure}

We finally present a merged cross-identified new general catalog of CAROs
in the SMC and MB, with 2741 entries. Figure~\ref{fig:objtypes} shows the
angular positions of the six grouped object classes:
clusters (C, CA, CC), emissionless associations (A, AC),
clusters and associations with emission (CN, NC, AN, NA), SNRs, ENs,
and TDGs (see Table~\ref{tab:census} and \citetalias{2008MNRAS.389..678B}
for a description of these classes).
ENs are emission nebulae without any obvious association or cluster.
The star-forming regions in the SMC main body, Wing and Bridge are evident.
Figure~\ref{fig:objtypes} illustrates a good definition of the SMC halo
clusters owing to their increase in number.
Both the star formation burst throughout
the main body, Wing and Bridge, and the inflated halo are part of the
same phenomenon:
the SMC disruption in the last (or last few) encounters with the LMC
\citep[e.g.][\citetalias{2008MNRAS.389..678B} and \citetalias{2019AJ....157...12B}]{2016AA...591...11D}.
The new OGLE-IV and SMASH clusters in the SMC halo and Bridge are important
to be studied in detail to disentangle Bridge young clusters from tidally
stripped halo or disk clusters in the Clouds.

Figure~\ref{fig:corclasses} shows the angular positions of the catalog objects,
color-coded by their relations to the \citetalias{2018ApJ...853..104B} sample,
where the ``I'', ``N'', ``E'', ``U'' and ``R'' classes are as 
defined in the previous section.

% Table 2
\begin{deluxetable*}{l@{}r@{}r@{}c@{}c@{}c@{}c@{}c@{}c@{}ccl@{}}
  \rotate
  \tablecaption{The general SMC/Bridge Catalog$^1$\label{tab:catalog}}
  \tablewidth{0.8pt}
    \scalefont{1.0}
\tablehead{
  \colhead{Designations} & \colhead{J2000 R. A.} & \colhead{J2000 Dec.} & \colhead{Type$^2$} & \colhead{$D^3$} & \colhead{$d^4$} & \colhead{Cl.$^5$} & \colhead{log(Age)} & \colhead{[M/H]} & \colhead{ref1$^{6,\dagger}$} & \colhead{ref2$^{7,\ddagger}$}
%  & \phantom{-}\phantom{-}\colhead{Comments$^7$} & \\
  & \colhead{Comments$^8$} \\
\colhead{}             & \colhead{(hh:mm:ss.s)} & \colhead{(Deg:$\arcmin$:$\arcsec$)} & \colhead{} & \colhead{($\arcmin$)} & \colhead{($\arcmin$)} & \colhead{} & \colhead{} & \colhead{} & \colhead{}  & \colhead{} & \colhead{} \\
}
\decimalcolnumbers
\startdata
%AM 3, ESO28SC4, OGLS 315         & 23:48:59.3 & -72:56:46 & C & 0.90 & 0.90 & U & \\
%L1, ESO28SC8, OGLS 313           &  0:03:54.6 & -73:28:16 & C & 4.60 & 4.60 & U & Globular Cluster? \\
%L2, OGLS 312, OGLS 328           &  0:12:56.9 & -73:29:28 & C & 1.20 & 1.20 & U & \\
%OGLS 264                         &  0:18:22.1 & -71:27:02 & C & 0.60 & 0.60 & U & \\
%L3,ESO28SC13, OGLS 323, OGLS 327 &  0:18:25.2 & -74:19:05 & C & 1.00 & 1.00 & U & \\
AM 3,ESO28SC4,OGLS 315           & 23:48:59.3 & -72:56:46 & C & 0.90 & 0.90 & U & 9.72 & -0.98 & PGC+14,DKB+14 & DKB+16 & \\
L1,ESO28SC8,OGLS 313             &  0:03:54.6 & -73:28:16 & C & 4.60 & 4.60 & U & 9.88 & -1.04 & GGS08,DKB+16  & PGC+15 & Globular Cluster ? \\
L2,OGLS 312,OGLS 328             &  0:12:56.9 & -73:29:28 & C & 1.20 & 1.20 & U & 9.6  & -1.58 & DKB+14        & DKB+16 & \\
OGLS 264                         &  0:18:22.1 & -71:27:02 & C & 0.60 & 0.60 & U & -    & -     &               &        & \\
L3,ESO28SC13,OGLS 323,OGLS 327   &  0:18:25.2 & -74:19:05 & C & 1.00 & 1.00 & U & 8.99 & -0.65 & PGC+14,DKB+14 & DKB+16 & \\
...                              &  ...       & ...     & ... & ... & ... & ... & ...  & ...   & ...           & ...    & ... \\
\enddata
\tablecomments{$^1$Only the first five entries are listed here; the full table
is available online in electronic format. The marked columns correspond to:
  $^2$Type of object as defined in Table~\ref{tab:census};
  $^3$Major angular size;
  $^4$Minor angular size;
  $^5$Class of correlation with the \citetalias{2018ApJ...853..104B} catalog;
  $^6$References for age;
  $^7$References for metallicity; and
  $^8$comments and hierarchical relation to other catalog objects (see text).\\
$^{\dagger}$From newest to oldest, the references for the ages stand for: 
%\citet{2018ApJ...852...68C,2016MNRAS.460..383P,2016AA...591...11D,2015MNRAS.450..552P,2015BAAA...57..102P,2015MNRAS.453.3190B,2014MNRAS.437.2005M,2014MNRAS.445.2302P,2014AA...561A.106D,2014AJ....147...71P,2012MNRAS.425.3085P,2010AA...517A..50G,2008MNRAS.391..915B,2008AJ....135.1106G,2008AJ....136.1703G,2008ApJ...681..290S,2007AJ....133...44S,2007ApJ...664..322R,2006AA...452..179C,2005AA...440..111P,2005AJ....129.2701R,2007ApJ...665L.109C,1999AcA....49..157P,1995MNRAS.276.1155S,1995AA...298...87G,1992AJ....103.1234G,1991AJ....101..911D,1990AJ....100..663G,1981AAS...46...79V,2018ApJ...853..104B}
\citet[BGB+18]{2018ApJ...853..104B};
\citet[CJK+18]{2018ApJ...852...68C};
\citet[DKB+16]{2016AA...591...11D};
\citet[PIR+16]{2016MNRAS.460..383P};
\citet[BSB+15]{2015MNRAS.453.3190B};
\citet[PCG+15]{2015BAAA...57..102P};
\citet[PdGR+15]{2015MNRAS.450..552P};
\citet[DKB+14]{2014AA...561A.106D};
\citet[MPS14]{2014MNRAS.437.2005M};
\citet[PGC+14]{2014AJ....147...71P};
\citet[P14]{2014MNRAS.445.2302P};
\citet[PB12]{2012MNRAS.425.3085P};
\citet[GGK10]{2010AA...517A..50G};
\citet[BSS08]{2008MNRAS.391..915B};
\citet[GGG+08]{2008AJ....135.1106G};
\citet[GGS+08]{2008AJ....136.1703G};
\citet[SGD+08]{2008ApJ...681..290S};
\citet[SSN+07]{2007AJ....133...44S};
\citet[RGB+07]{2007ApJ...664..322R};
\citet[CVH+06]{2006AA...452..179C};
\citet[PSC+05]{2005AA...440..111P};
\citet[RZ05]{2005AJ....129.2701R};
\citet[CSS+07]{2007ApJ...665L.109C};
\citet[PU99]{1999AcA....49..157P};
\citet[GCB+95]{1995AA...298...87G};
\citet[SBC+95]{1995MNRAS.276.1155S};
\citet[GDK92]{1992AJ....103.1234G};
\citet[DGI+91]{1991AJ....101..911D};
\citet[GDK+90]{1990AJ....100..663G}; and
\citet[vdB81]{1981AAS...46...79V}.\\
$^{\ddagger}$For the metallicities, besides the aforementioned references,
we adopted:
%\citet{2015AJ....149..154P,2009AJ....138..517P,1998AJ....115.1934D,1999AA...345..430H,2010MNRAS.403.1156R,2007AJ....133...44S,2007ApJ...664..322R,2001AJ....122..220C,2001ApJ...562..303D,1999ApJ...511..225A,1998AJ....116.2395M,2014AJ....147...71P,2016AA...591...11D,2014AA...561A.106D,2017AA...602A..89P,2015MNRAS.450..552P,2011MNRAS.416L..89P,2011MNRAS.418L..69P,2012ApJ...756L..32P,2008MNRAS.391..915B,2015MNRAS.453.3190B,2011MNRAS.417.1559P,2007MNRAS.377..300P,2005MNRAS.358.1215P,2001MNRAS.325..792P,2010MNRAS.403..797G,2010AA...520A..85D,2005AA...440..111P,1998AA...332...19D,1999IAUS..190..446D,1996AJ....112.2004A,1992ApJS...82..489M,1984ApJ...286..517R,1984ApJS...55...45Z,1989rdmc.conf...45S,1986AA...156..261B}.
\citet[PPV17]{2017AA...602A..89P};
\citet[PGC+15]{2015AJ....149..154P};
\citet[P12]{2012ApJ...756L..32P};
\citet[P11a]{2011MNRAS.416L..89P};
\citet[P11b]{2011MNRAS.418L..69P};
\citet[PCB+11]{2011MNRAS.417.1559P};
\citet[DCB+10]{2010AA...520A..85D};
\citet[GDC10]{2010MNRAS.403..797G};
\citet[RKG10]{2010MNRAS.403.1156R};
\citet[PGG+09]{2009AJ....138..517P};
\citet[PSG+07]{2007MNRAS.377..300P};
\citet[PSG+05]{2005MNRAS.358.1215P};
\citet[PSC+01]{2001MNRAS.325..792P};
\citet[CSP+01]{2001AJ....122..220C};
\citet[DWH+01]{2001ApJ...562..303D};
\citet[AS99]{1999ApJ...511..225A};
\citet[D99]{1999IAUS..190..446D};
\citet[H99]{1999AA...345..430H};
\citet[dFPBI98]{1998AA...332...19D};
\citet[DH98]{1998AJ....115.1934D};
\citet[MSF98]{1998AJ....116.2395M};
\citet[ALA+96]{1996AJ....112.2004A};
\citet[MJdC92]{1992ApJS...82..489M};
\citet[SR89]{1989rdmc.conf...45S};
\citet[BDP86]{1986AA...156..261B}
\citet[RdCM84]{1984ApJ...286..517R}; and
\citet[ZW84]{1984ApJS...55...45Z}.}
\end{deluxetable*}

Table~\ref{tab:census} gives an updated census of the object classes and
their counts,
including different classes of correlation with the
\citetalias{2018ApJ...853..104B} catalog.
These classifications allow to peer and discriminate the new catalog
contents, and have been used in several studies of the Clouds \citep[][and
references therein]{2008MNRAS.389..678B}.
The present general catalog is a factor $\sim$2 larger than its
\citetalias{2008MNRAS.389..678B} counterpart.

% Table 3
\begin{deluxetable*}{ll|l|rrrrr|r}
  \rotate
  \tablecaption{Updated Census of the SMC and Bridge extended objects by object
    class
and correlation with the \citetalias{2018ApJ...853..104B} Catalog.\label{tab:census}}
\tablewidth{0pt}
\tablehead{
\colhead{Object Class} & \colhead{Characteristics} & \colhead{Description} &
\colhead{I} & \colhead{N} & \colhead{E} & \colhead{U} & \colhead{R} &
\colhead{total}
}
\decimalcolnumbers
\startdata
C   & star cluster & resolved star cluster &
  0 &    1 &   58 &  529 &   38 &   626 \\
CA  & poor cluster transition to small assoc. & structure looser than clusters &
  0 &    0 &   10 &  133 &   13 &   156 \\
A   & association & --- &
960 &  207 &   21 &  210 &    9 &  1407 \\
AC  & small association, looser than clusters & association character dominates &
  0 &    1 &    3 &   62 &    2 &    68 \\
CC  & cluster candidate & non-resolved cluster &
  0 &    0 &    0 &   39 &    2 &    41 \\
NC  & cluster in emission & cluster in nebula, dominated by gas emission & 
  0 &    0 &    5 &  122 &    5 &   132 \\
CN  & cluster with some emission & cluster signature, dominated by stars &
  0 &    0 &    1 &   22 &    2 &    25 \\
NA  & association in emission &  dominated by gas emission (mostly HII regions) &
  0 &    2 &   14 &  166 &   17 &   199 \\
AN  & associations with some emission & dominated by stars  &
  1 &    3 &    7 &   33 &    8 &    52 \\
EN  & Nebula without association or cluster & ---  &
  0 &    0 &    0 &    6 &    0 &     6 \\ 
SNR & supernova remnant & Type II SNRs trace star forming regions &
  0 &    0 &    0 &   26 &    0 &    26 \\
TDG & tidal dwarf galaxy & Concentrations of objects in the Bridge &
  0 &    0 &    0 &    3 &    0 &     3 \\
\hline
total & & &
961 &  214 &  119 & 1351 &   96 &  2741 \\
\enddata
\end{deluxetable*}

\section{Metallicities and ages}\label{sec:AgeMet}

% Fig. 3
\begin{figure*}[ht!]
%\plotone{SMCcatFig03.pdf}
\plotone{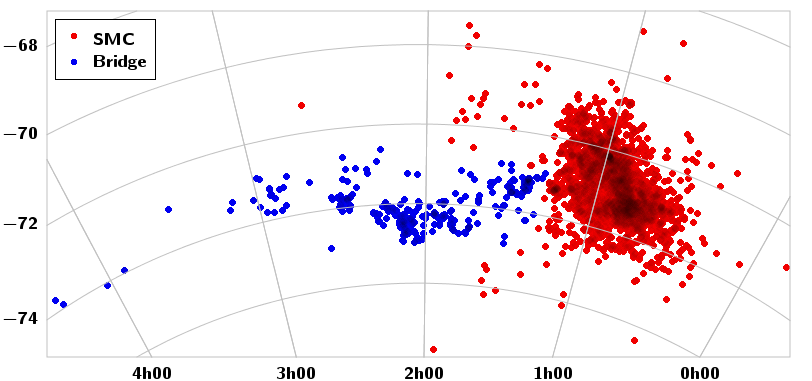}
%  !!!! Escrever caption !!!
\caption{
  Spatial distribution of the ages and metallicities for the objects
  in the catalog.
  Objects in grey have no age or metallicity value in our catalog.
  \label{fig:agemetmaps}}
\end{figure*}

% Fig. 4
\begin{figure*}[ht!]
%\plotone{SMCcat_Fig4.png}
%\plotone{SMCcat_Fig4.pdf}
\plotone{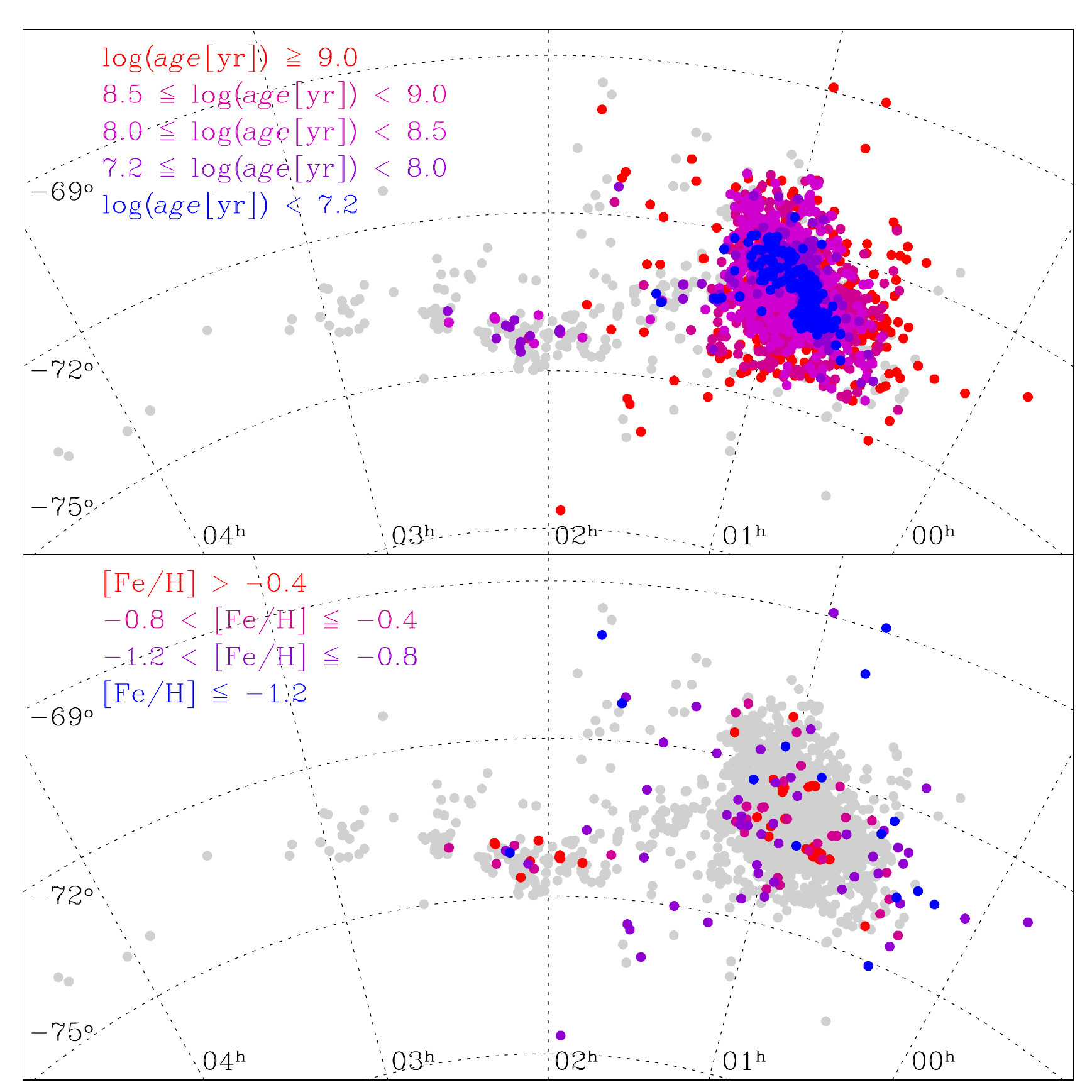}
\caption{
  Separation of the SMC objects into Main Body (red) and Bridge + Wing
  (blue) samples.
  \label{fig:bridge-selec}}
\end{figure*}

% Fig. 5
\begin{figure}[ht!]
%\plotone{FEH_Hist_rice-11_Full.pdf}
\plotone{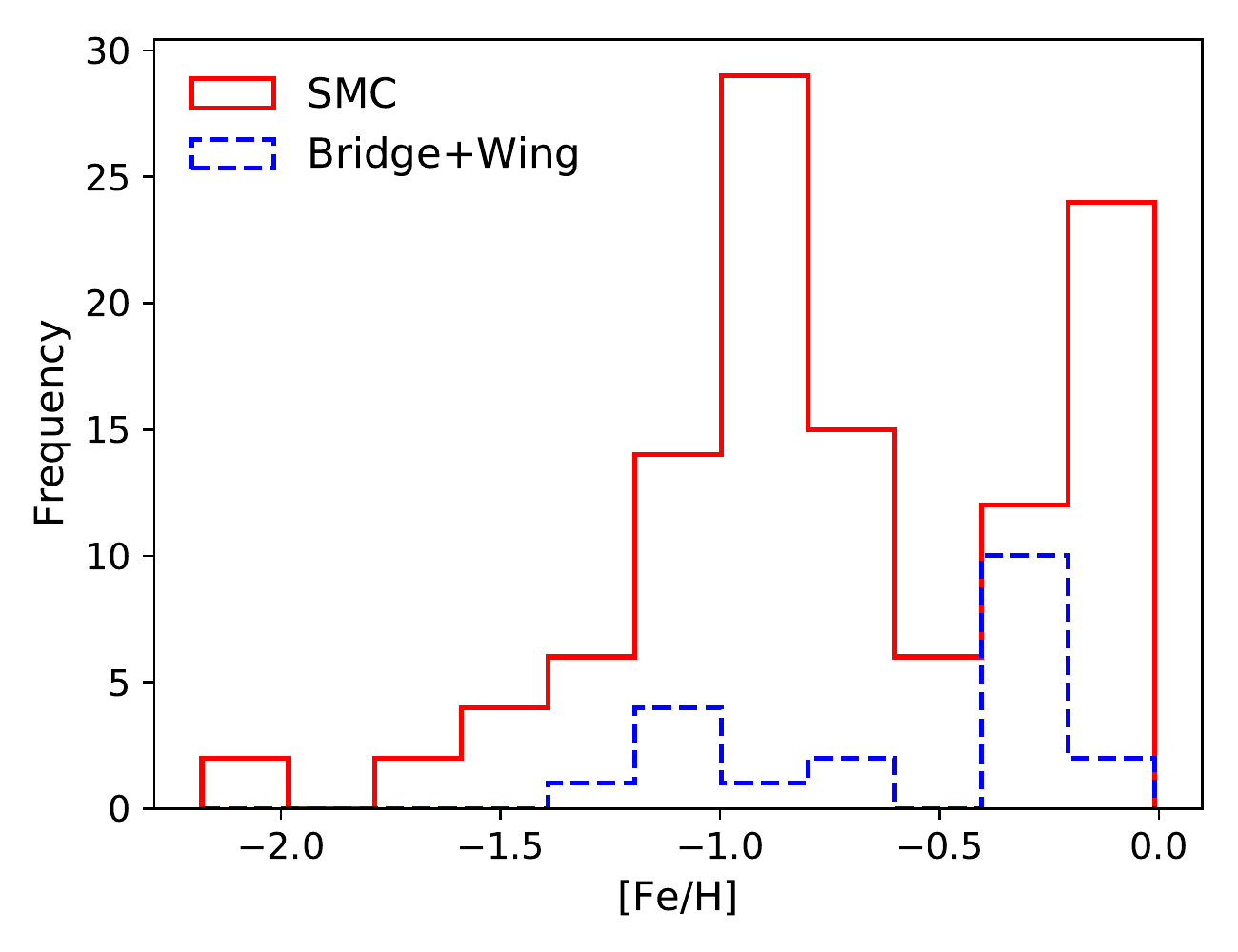}
\caption{Histogram of metallicities in bins of 0.2 dex,
subdivided in SMC and Bridge + Wing clusters.\label{fig:met-hist}}
\end{figure}

% Fig. 6
\begin{figure}[ht!]
%\plotone{Age_Hist.pdf}
\plotone{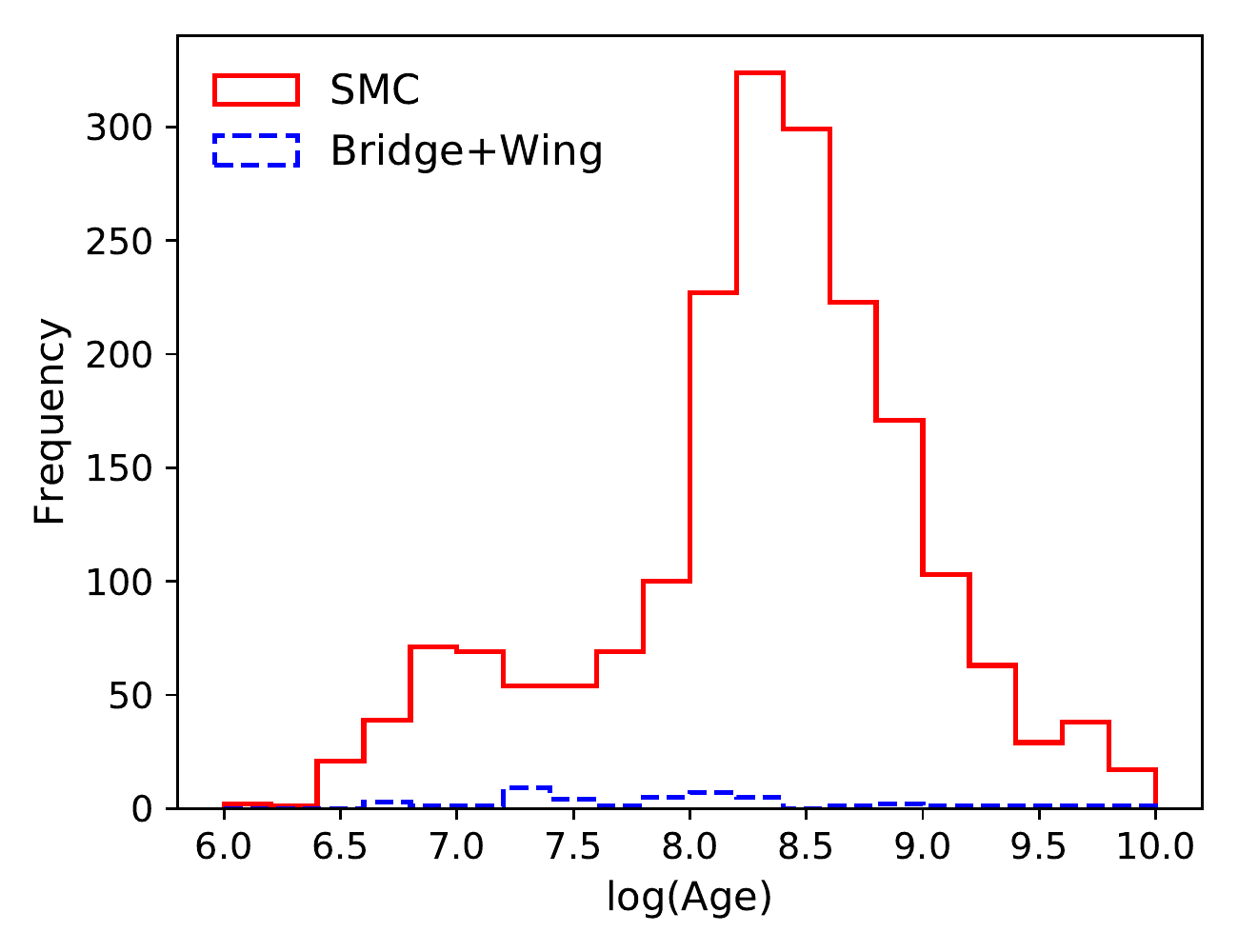}
\caption{Histogram of ages,
subdivided in SMC and Bridge + Wing clusters.\label{fig:age-hist}}
\end{figure}

% Fig. 7
%\begin{figure*}[ht!]
\begin{figure}[ht!]
%\includegraphics[width=1.0\textwidth]{amr+bridge.pdf}
%\plotone{amr+bridge.pdf}
\plotone{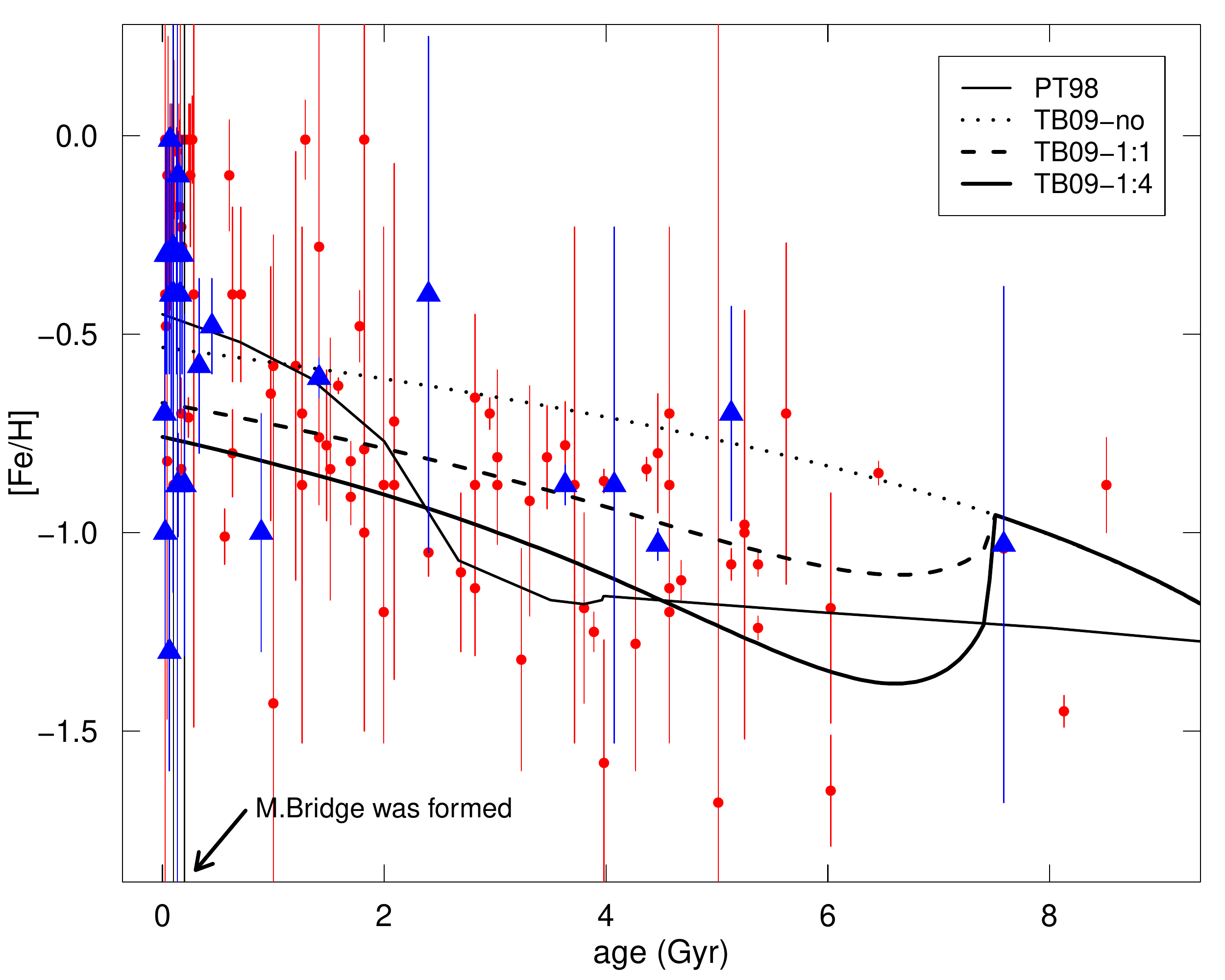}
\caption{Age-metallicity distribution of SMC CAROs with information
compiled in Table~\ref{tab:catalog}.
The error bars represent the errors given by the respective authors.
The Wing/Bridge clusters are shown as blue triangles.
The formation time of the Magellanic Bridge is indicated by vertical lines
at 100-200~Myr \citep[e. g.][]{2018ApJ...864...55Z}.
The burst model of \citet{1998MNRAS.299..535P} and the merger models of
\citet{2009ApJ...700L..69T} are shown to illustrate how a single model is unable
to reproduce the age-metallicity distribution of SMC CAROs.\label{fig:MetAgerel}}
%\end{figure*}
\end{figure}

% Fig. 8
%\begin{figure*}[ht!]
\begin{figure}[ht!]
%\includegraphics[width=1.0\textwidth]{grad-all.pdf}
%\plotone{grad-all.pdf}
%\plotone{grad-all+sigma.pdf}
\plotone{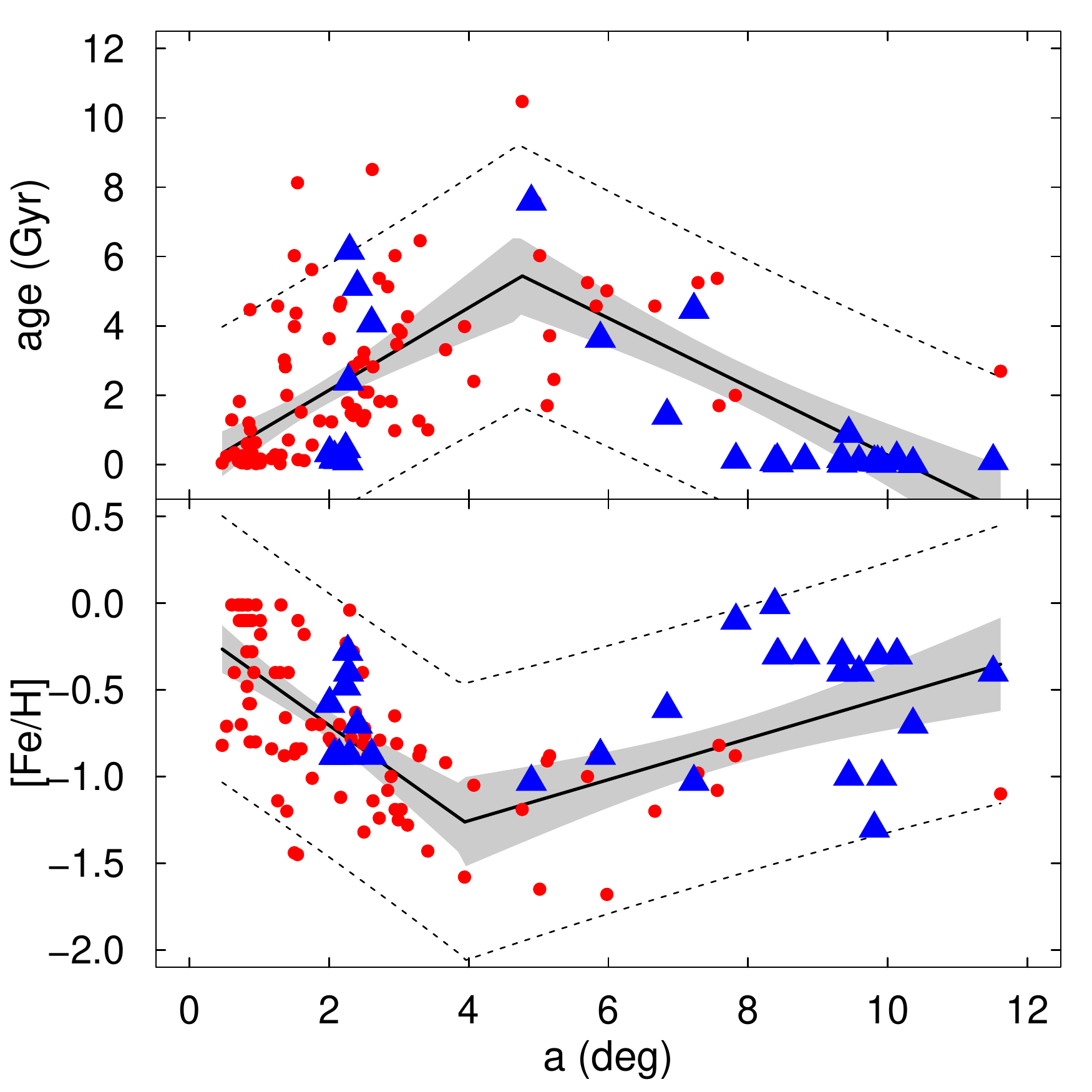}
\caption{Age and metallicity of SMC clusters from the literature
 compilation in Table~\ref{tab:catalog}, as a function of the projected distance from
 the SMC centre. The distance is the semi-major axis of the ellipse surrounding
 the SMC main body as defined by \citet{2007MNRAS.377..300P} and used by
 \citet{2014AA...561A.106D,2016AA...591...11D,2016AJ....152...58P}.
 Bridge clusters are highlighted as blue triangles.
 The shaded grey area represents the 95\% confidence interval of
   the fitted parameters. The dashed lines represent the 95\% prediction
   interval of the points.
 We fitted all points with a linear regression with a breakpoint using
 the R package ``segmented'' \citep{2003Muggeo}
 to highlight the inversion
 in both gradients at about $a \sim 4-5^{\circ}$.\label{fig:MetAgegrad}}
%\end{figure*}
\end{figure}

Table~\ref{tab:catalog} also includes in columns~8 to 11 a compilation
of ages and metallicities from the literature, together with corresponding
references and abbreviations.
Since they include the results of \citetalias{2018ApJ...853..104B}, it is the
most complete sample available.

For the cases where more than one reliable age determination was available, we took the
average in log(Age).
HW41, B112 and HW81 have double structure, so we do not include the single
object ages.
Since H86-106 may have two components, we do not include the age from the
literature either.

For the metalicities, we selected the most reliable determinations,
favouring, in this order: calcium triplet and other high resolution
spectroscopic determinations from individual (giant) stars,
isochrone fitting in the CMD, in a few cases
integrated spectroscopy, and 
if no other metallicity determinations were available --
the \citet{1986AA...156..261B} integrated photometry values.
For each object, the age and metallicity references
are given in columns 10 and 11 of Table~\ref{tab:catalog}.
% as abbreviations, which are explained in the table notes.

In Figure~\ref{fig:agemetmaps}, the positions of the catalog
objects with literature age and/or metallicity determinations can be seen,
color-coded by different ranges of these parameters.
As an aid to see where they are located with respect to the LMC or
Bridge, we show the objects without known age or metallicity in grey.

Figure~\ref{fig:bridge-selec} shows the CAROs
identified as Bridge/Wing. For the sake of simplicity, we considered as belonging
to the Bridge the objects between $1^{h}20^{m} < \rm{RA} < 4^{h}30^{m}$ and
$-75^{\circ}< \rm{Dec} <-72^{\circ}$, thus including the SMC Wing.

In Figure~\ref{fig:met-hist} is shown the histogram of metallicities, separated
as belonging to the SMC or Bridge+Wing. Among the 2741 entries, 626 are
confirmed clusters, and metallicity derivations are mostly based on some of these
clusters that amount to 117, plus a few associations. Therefore, only 134 clusters and associations (5\%) have spectroscopic
metallicities in the literature, whereas ages are available for 75\% of them.

Testing statistical techniques to select the optimal bin width, we adopted the
square root of the number of clusters, obtaining 11 bins of 0.2~dex. Of the 26
objects presented in the most metal-rich bin \citep[$-0.2<\rm{[Fe/H]}<0.0$, all
from][]{2017AA...602A..89P}, twelve objects have $\rm{[Fe/H]}=-0.01$, which is
the upper limit of the parameter space explored by them.
If these objects were excluded from the analysis, this most metal-rich bin
would then drop by half. The metallicity distribution presents a peak at
$\rm{[Fe/H]} \sim-0.8$ to $-1.0$. This is in general terms in agreement with the
recent literature on the stellar populations of the SMC: a mean metallicity of
$\rm{[Fe/H]}\sim-0.7$ is identified for the young populations
\citep{2018MNRAS.477..421K};
and metallicities of $\rm{[Fe/H]}\sim-0.8$ to $-1.0$ are
assumed for red giant stars \citep{2016ARA&A..54..363D}.

\citet{2015AJ....149..154P}
found a metallicity distribution of clusters based on CaII triplet spectroscopic
metallicity ranging from $-1.4< \rm{[Fe/H]} <-0.4$ with a possible bimodality.
Nevertheless, a non-negligible number of clusters are more metal-poor than
$\rm{[Fe/H]}<-1.5$, and other $\sim 50$ ones are more metal-rich than
$\rm{[Fe/H]}>-0.5$. Recently a photometric metallicity map of the SMC was
presented by \citet{2018MNRAS.475.4279C}, showing no field stars with
$\rm{[Fe/H]}<-1.2$, and $\rm{[Fe/H]}>-0.7$. Since this
technique is limited to find
more metal-poor stars, spectroscopy is required. \citet{2016AJ....152...58P}
found a distribution of metallicities for field stars ranging from
$-2.4 < \rm{[Fe/H]} < -0.2$, based on CaII triplet spectroscopy.
Even considering that
clusters could be captured from the LMC or from the Galaxy, and the
low-metallicity and high-metallicity ones could be explained in that way, we
would suggest that such clusters should be reanalyzed with high-resolution
spectroscopy.

In Figure~\ref{fig:age-hist} the histogram of ages is shown, with a fixed
bin width of 0.2 in log(Age). It is interesting to note that a large
number of them, amounting to 225 objects (of a total of 2019 objects
or $75\%$ of the sample), are older than 1\,Gyr, which makes this sample
of great interest for studies of the early formation of the SMC.

The age histogram suggests a major event of star formation at around
180\,Myr, as could have been triggered by an encounter between
the SMC and the LMC. This is the estimated age
of the Magellanic Bridge based on dynamical studies of the last
encounter between LMC-SMC \citep[e.g.][]{2018ApJ...864...55Z}. Looking at
the histogram in blue where only Bridge objects are represented, it is clear
that the star formation was quiescent until about $\sim 150$\,Myr ago.
However, the decay for $t>500\,{\rm Myr}$ clusters can be eroded by cluster
dissolution effects, as in the Milky Way \citep{2011MNRAS.415.2827B}. Old
low mass clusters are either too faint or have mostly been dissolved
\citep{2012MNRAS.423.1390B}.

The age-metallicity relation (AMR) of the SMC CAROs has been subject
of considerable investigation. \citet{2015AJ....149..154P} have found that
even with a homogeneous sample, there is an intrinsic metallicity dispersion at
a given age, concluding that no single chemical evolution model can describe
the evolution of the SMC.
\citet{2014AA...561A.106D,2016AA...591...11D} proposed to conduct
this study by splitting the SMC into four groups related to the SMC-LMC-MW
tidal interactions, namely, the main body and three external groups
that are being stripped out from the main body: Wing/Bridge, counter-bridge,
and west halo. They pointed out the need of a homogeneous sample of ages
and metallicities to make any reliable conclusions.
Although we could not find a dip in metallicity in the AMR of the Wing/Bridge
clusters in our sample, probably due to the highly heterogeneous sample,
as can be seen in Figure~\ref{fig:MetAgerel},
we were still able to recover the inversion in the age and metallicity radial
gradients found in the aforementioned works, as shown in
Figure~\ref{fig:MetAgegrad}.
We highlight the Wing/Bridge clusters and conclude that the inverted gradient
out of $a \gtrsim 4^{\circ}$ seems to be dominated by Wing/Bridge clusters.
A further detailed study with a homogeneous sample will be carried out in
a future work.

\section{Distant Clusters and New outer limits for the Magellanic System}\label{sec:landmarks}

% Fig. 9
\begin{figure*}[ht!]
%  \plotone{SMCcatFig3.eps}
%  \includegraphics[width=1.0\textwidth]{SMCcatFig3.eps}
%  \includegraphics[width=1.0\textwidth]{SMCcatFig3.pdf}
  \includegraphics[width=1.0\textwidth]{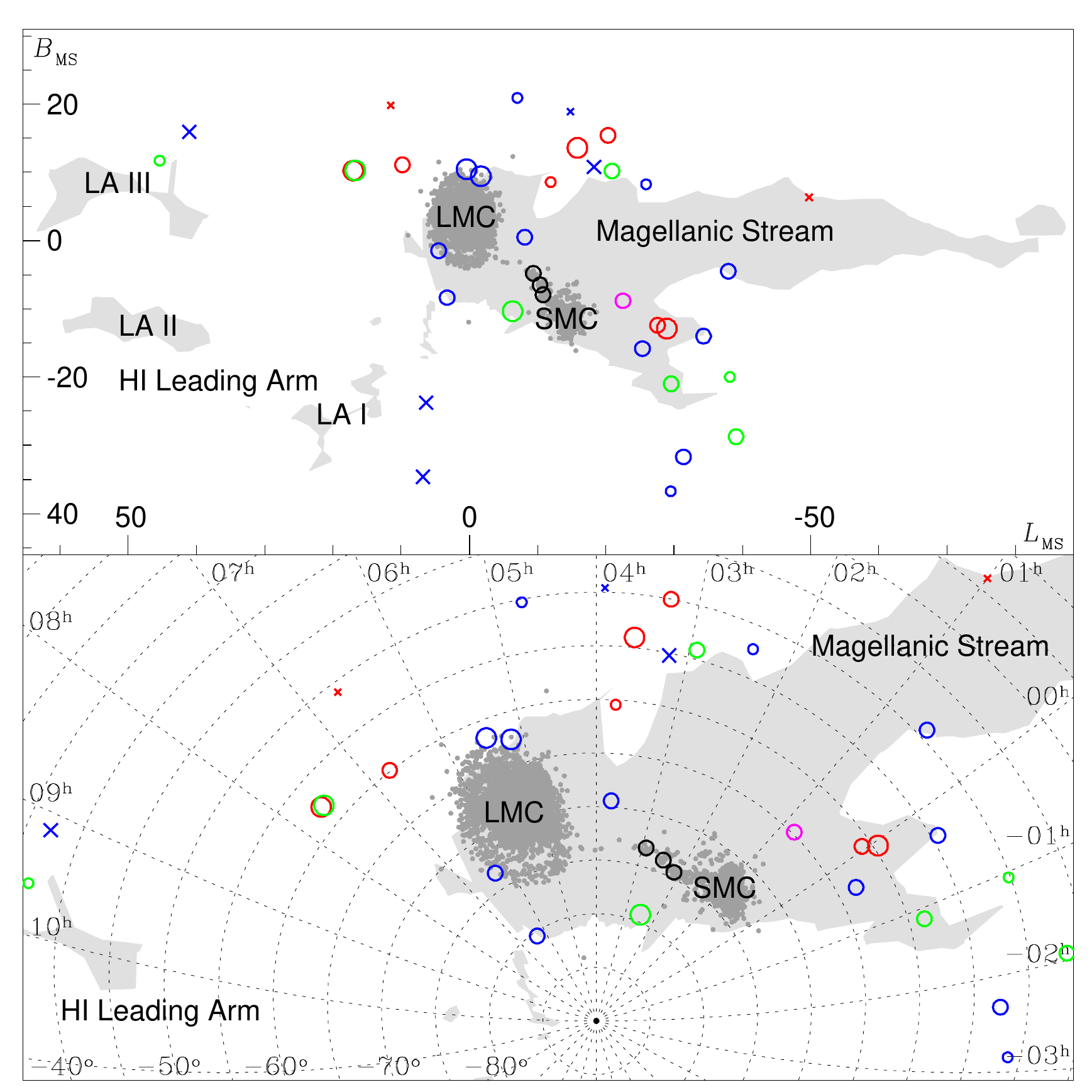}
\caption{EMS angular distribution.
Grey points: the present SMC/MB catalog, and LMC CAROs from \citetalias{2008MNRAS.389..678B};
blue: ultra-faint star clusters (UFCs); red: ultra-faint dwarf galaxies (UFGs),
green: faint dwarf galaxies (FGs); magenta: tidal debris SMCNOD;
open black circles: tidal dwarf galaxies from \citetalias{1995ApJS..101...41B}.
red crosses: dwarf spheroidals; blue crosses: MW globular clusters;
large symbols: objects up to 40~kpc from the Sun,
intermediately sized symbols: objects between 40~kpc and 90~kpc,
small symbols: objects from 90~kpc distance on;
Shaded regions in light grey: The Magellanic Stream and the HI Leading arm.
The contours of these gas structures were extracted by one of us (B. D.)
from Figure~8 of \citet{2008ApJ...679..432N} to represent the gas
distribution regions without any information on gas density or velocity.
LA I, LA II, and LA III are the Leading Arm complexes I, II and III
identified by \citet{2008ApJ...679..432N}.
Top panel: The EMS object sample in relation to the two gas structures.
Here, we use the Magellanic Stream Coordinate System as defined by
\citet{2008ApJ...679..432N};
Bottom panel: Enlargement of the region, where the EMS object sample
lies.\label{fig:landmarks}}
\end{figure*}

Discoveries of ultra-faint star clusters (UFC), ultra-faint (UFG) and
faint (FG) dwarf galaxies around the Clouds have been mostly
carried out with
the Dark Energy Survey \citep[DES,][]{2015ApJ...813..109D}.
Deep photometric, spectroscopic, kinematical and dynamical follow-ups
probed them further \citep[e. g.][]{2018ApJ...852...68C}.
The UFCs can be used to establish new landmarks and frontiers for an EMS.
The catalog of the SMC/MB objects must cope with that
involving the MW, LMC and SMC potentials.
The MW has certainly captured clusters which originated in the LMC and SMC,
and some of their satellite galaxies.
The relevant FGs and UFGs are projected around the Clouds at various
heliocentric distances, in front or behind them.
They are or were LMC satellites
\citep[][and references therein; \citealt{2018ApJ...857..145L}]{2018arXiv180902259J}.
The UFCs Pic~I and Phe~II, as well as the UFG Grus~I present
tidal substructures pointing to the LMC. The UFGs Hor~I, Car~II, Car~III
and Grus~I have been suggested to be related to the LMC,
while Tuc~II and Tuc~IV might be related to the SMC, together with the
UFCs DES~1 and Eri~III \citep{2018ApJ...852...68C}.
The UFG Hydrus~I probably originated together with the LMC and migrated
to the MW halo \citep{2018MNRAS.479..534K}, while Grus~I was probably captured by
the MW on the MC far side. Figure~\ref{fig:landmarks} shows the angular
distribution of the objects in Table~\ref{tab:landmarks} (Appendix~\ref{AppTable4}),
from the east in the LMC Leading Arm to far west of the SMC, trailing the MC.
The present discussion deals with the entire EMS, to be joined by the
updated LMC catalog in a forthcoming study.

The LMC UFG neighbors, whether satellites, captures, dissolving or comoving,
can provide constraints
on the formation and hierarchical evolution of galaxies
\citep{2017MNRAS.472.1060D}.
Table~\ref{tab:landmarks} (Appendix~\ref{AppTable4}) gives 27 objects, their characterizations
and references, containing UFCs, FGs, UFGs, tidal galaxies and/or tidal debris.
Several scenarios can operate:
{\it (i)} co-movers with the Clouds in the Vast Polar Structure
\citep[VPO,][]{2014ApJ...790...74P},
{\it (ii)} satellites formed in or around the Clouds and eventually captured by the MW,
{\it (iii)} objects originated in the MC and captured by the MW, and
{\it (iv)} plain clusters originated in the LMC or SMC that remain captive.
In the last column of Table~\ref{tab:landmarks} we also show diagnostics
on the object nature
according to each paper, based on position, age, metallicity,
total absolute magnitude, dark matter content, and/or orbits.
In some cases we complemented them.
The objects are contained in an area with angular separation
$<40~{\rm degrees}$ from the LMC and heliocentric
distances $15~{\rm kpc} < d_{\odot} < 130~{\rm kpc}$.
It includes a considerable MW halo slice and engulfs the possibility
of scattered objects with
$d_{\odot}$ a factor $\sim$2 of the SMC and LMC distances of 59 and 49~kpc,
respectively, derived from Cepheids \citep{2018A&A...620...99G}.

Figure~\ref{fig:landmarks} shows the objects of Table~\ref{tab:landmarks}
and suggests relationships within the EMS.
The tidal dwarf galaxies BS~I, BS~II and BS~III \citepalias{1995ApJS..101...41B}
in the Bridge
may evolve to SMCNOD-like \citep[SMC Northern Over-Density, ][]{2017MNRAS.468.1349P} overdensities,
which are long-lived tidal debris.
While the BS TDGs are gas-rich with an essentially young stellar content
\citepalias{2015MNRAS.453.3190B}
SMCNOD has an intermediate age population.
They may be different evolutionary stages of a process creating tidal dwarf
galaxies (\citetalias{2015MNRAS.453.3190B} and references therein).
SMCNOD on the SMC side, as well as Antlia~II \citep{2019MNRAS.488.2743T}
which is probably related to the LMC Leading Arm
may be evolved examples of TDGs, or alternatively, tidal debris.
On the other hand, Ant~II may represent one of the most diffuse
genuine early galaxies \citep{2019MNRAS.488.2743T}.

Objects related to the LMC or SMC are not restricted to the
area studied here, which is expected to englobe an EMS.
\citet{2018ApJ...867...19K} found that Hydrus~I, Car~II, Car~III and Hor~I,
which are within this area, have kinematics consistent with the LMC.
Furthermore, Hydra~II (outside the area), and especially Dra~II
(far outside) may be kinematically related to the LMC, and deserve more analysis
in the future.
Orbit calculations can indicate complex interaction scenarios,
e. g. for Tuc~III,
an UFG with a stream and projected near the SMC.
It appears to have endured a close encounter with the LMC at 75\,Myr ago
\citep{2018MNRAS.481.3148E}, when it was cast
into the MW halo, and is in dissolution.
Table~\ref{tab:landmarks} indicates the objects that have kinematical
(radial velocity or proper motion) or dynamical (orbital) information.
Many of the UFGs and UFCs have kinematical/dynamical data,
and in general they support a physical connection with the Clouds.

The $\Lambda$ Cold Dark Matter theories predict that
the halos of galaxies like the LMC should include
about 50 dwarf companions \citep{2017MNRAS.472.1060D}.
Several of them appear to have been detected (Table~\ref{tab:landmarks}).
Despite the massive search efforts, there is a deficit of companion
galaxies, while initially classified as UFG candidates
turned out to be UFCs, as shown by follow-up studies,
such as Eri~III \citep{2018arXiv180902259J}, Pic~I
and probably Phe~II \citep{2018ApJ...852...68C}. Table~\ref{tab:landmarks}
contains 13 UFCs, 7 FGs and 7 UFGs,
when placing the limit between FG and UFG/UFCs at $M_V=-3.5$.
LMC satellites are still missing \citep[][present study]{2017MNRAS.472.1060D}.
Possibilities are:
{\it (i)} dwarf galaxy dissolutions have been frequent, as the MC plunged into
the MW halo;
{\it (ii)} fainter galaxies will be discovered,
especially UFGs or extended low density FGs like Ant~II;
{\it (iii)} or alternatively, some changes are needed in early Universe models
\citep{2017MNRAS.472.1060D}.

Two dwarf spheroidals and five MW halo GCs are located within the
area studied here.
Orbit calculations \citep{2018A&A...616A..12G} showed that Sculptor
resides between an apocenter of 111.8~kpc and a pericenter of 59.7~kpc,
and Carina between 107.5~kpc and 87.0~kpc.
The pericenter suggests that
Sculptor may have had interactions with the MC.
Orbits of the five MW halo GCs \citep{2019MNRAS.482.5138B} show that the apocenters of IC~4499
and NGC~1261 are smaller than 28~kpc,
suggesting early accretions in the hierarchical history of the Galaxy.
NGC~6101 with apocenter at 47~kpc may have interacted with the LMC.
NGC~6101, Pyxis [131.2~kpc, 26.3~kpc] and AM~1 [308.3~kpc, 98.8~kpc]
require mass models including the MC for reliable interpretations.

\citet{2016AcA....66..255S} discovered clusters in the outskirts of the
LMC, and \citet{2019MNRAS.484.2181T} derived parameters for them.
%  the UFCs Gaia~3, DES~4 and DES~5, with the OGLE-IV survey, and
%\citet{2019MNRAS.484.2181T} derived parameters for them.}
Thus we also include their OGLL designations in Table~\ref{tab:landmarks}.
They are projected near the edge of the LMC outer disk
(Figure~\ref{fig:landmarks}).
Gaia~3 has a compatible distance to the LMC (Table~\ref{tab:landmarks}),
while DES~4 and DES~5 are located somewhat in the LMC foreground,
suggesting capture by the MW potential.

The 13 UFCs as an ensemble (Table~\ref{tab:landmarks}, Figure~\ref{fig:landmarks})
suggests that the EMS is very extended, and that
most of them were formed in the Clouds and some others have migrated into the MW potential well.
However, the age-metallicity relations of the Clouds \citep{2013AJ....145...17P}
are not matched by the young age and low metallicity of OGLL~845 (Gaia~3),
which appears to have its origin in another dwarf galaxy.
Pic~I is an UFC whose orbit indicates it as an outer LMC member.
The MW and especially the LMC still require more realistic model potentials
\citep{2018MNRAS.481.3148E}.
\citet{2018ApJ...860...76H} recently argued that the Galactic
gravitational potential induces the dwarf line-of-sight velocity dispersion,
questioning the estimates of dark matter.
Table~\ref{tab:landmarks} gives hints, but to settle the
EMS benchmarks, more constraints are necessary,
both observational and theoretical.

\section{Concluding Remarks and Perspective on Future Work}\label{sec:conclusion}

We provide an updated census of star clusters, associations and other
related extended objects in the SMC and Magellanic Bridge.
Ten years have elapsed since the last general catalog effort,
and new cross-matches were necessary.
Interesting new clusters have been discovered in recent surveys,
such as OGLE-IV \citep{2017AcA....67..363S}
and SMASH \citep{2017ApJ...834L..14P} in the SMC halo and Bridge,
as well as
VMC central SMC bar clusters in the near IR \citep{2016MNRAS.460..383P}.
We communicate our own discovery of 64
clusters and candidates in the SMC and Bridge.

We also cross-identified these clusters and candidates with objects
from the SMC catalog by \citetalias{2018ApJ...853..104B}.
We clarified the issue of overestimated number of star clusters
\citep[see][]{2018MNRAS.478..784P}.
\citetalias{2018ApJ...853..104B} refer to their objects as star clusters, but most have low stellar
density and are in general diffuse and extended.
Consequently, we classified them as associations.
The census indicates that \citetalias{2018ApJ...853..104B} contributed with 1175 new SMC objects,
while 119 have previous counterparts.
Their sample contains essentially no faint clusters.
All in all, the present general catalog provides 2741 objects in the SMC
and Bridge (Table~\ref{tab:catalog}).

The present effort producing accurate coordinates and cross-matches
for the previous literature objects
will be useful for new cluster searches.
An example is by means of image
inspections by researchers and interested citizens, as organized by
SMASH\footnote{\url{https://www.zooniverse.org/projects/lcjohnso/local-group-cluster-search}}.
We point out that the present new clusters and candidates were not
systematically searched for, but were mostly found serendipitously while
analyzing the SMC and Bridge fields for previous objects.
The new updated, reliable coordinates and characterizations will be
particularly useful for observations, by minimizing uncertainties in crowded
cluster zones, or in the study of cluster pairs and multiplets.
It must be emphasized that the cluster center pointings in this paper provide
in general more accurate cluster coordinates than previous studies
because the latter searched for peaks in stellar or flux density distributions,
which as a rule have shifts owing to overcrowding and saturation effects.
The present catalog also contains ages and metallicities from the literature,
where available.

As a continuation of this work we will present a study of the LMC,
also starting off from \citetalias{2008MNRAS.389..678B} and adding new studies by means of
cross-identifications, in particular the
LMC analysis of \citet{2017ApJ...845...56B}.

A general SMC catalog must address the numerous UFCs, FGs and UFGs
surrounding the Clouds.
Table~\ref{tab:landmarks} compiles 27 such underluminous objects,
providing diagnostics for their nature, and the
probable relation to the Clouds or MW. Most of the FGs and UFGs are compatible
with being satellites of the Clouds,
while UFCs appear to have originated in the Clouds.

The present study was carried out within the framework of the ongoing project
VISCACHA \citep{2019MNRAS.484.5702M}.
This project employs the SOAR 4.1~m telescope with instrumental settings
determining the ages of massive and low mass MC clusters from their CMDs,
going deeper than the turn-off of old clusters in both Clouds,
dealing better with crowding than previous surveys, because of the
adaptive optics module SAM.
Currently, we are facing the curtain of low mass clusters in the SMC
\citep[e. g.][]{2012MNRAS.425.3085P}.
However, we have not yet unveiled them to show clusters with masses comparable
to open clusters in the MW, as the two clusters serendipitously found with HST
in a bar crowded field on the east side of the LMC \citep{1998MNRAS.295..860S}.
The present effort to gather all known clusters so far into a single
SMC and Bridge catalog with improved positions and other characteristics
will be particularly useful to probe the hidden population of faint clusters
in the Clouds.
In return, the VISCACHA results, i.~e. the properties of the observed stellar
clusters, will be implemented into the catalog.

\acknowledgments

The authors acknowledge support from the Brazilian Institutions CNPq,
FAPESP and FAPEMIG.
F.F.S.M. acknowledges FAPESP funding through the fellowship n$^{\rm o}$
2018/05535-3.
R.A.P.O. acknowledges the FAPESP PhD fellowship no. 2018/22181-0.
This study was financed in part by the Coordena\c{c}\~{a}o
de Aperfei\c{c}oamento de Pessoal de N\'{\i}vel Superior - Brasil
(CAPES) - Finance Code 001.  \\
This research has made use of ``Aladin sky atlas'' developed at CDS,
Strasbourg Observatory, France \citep{2000A&AS..143...33B,2014ASPC..485..277B}.
% \software{Aladin Sky Atlas \citep{2000A&AS..143...33B,2014ASPC..485..277B}
%           }
We thank an anonymous referee for interesting remarks.

\appendix
\restartappendixnumbering

\section{New clusters}
\label{AppNewClusters}

% Fig. 10 / A1
\begin{figure*}
\centering
\includegraphics[width=1.0\linewidth]{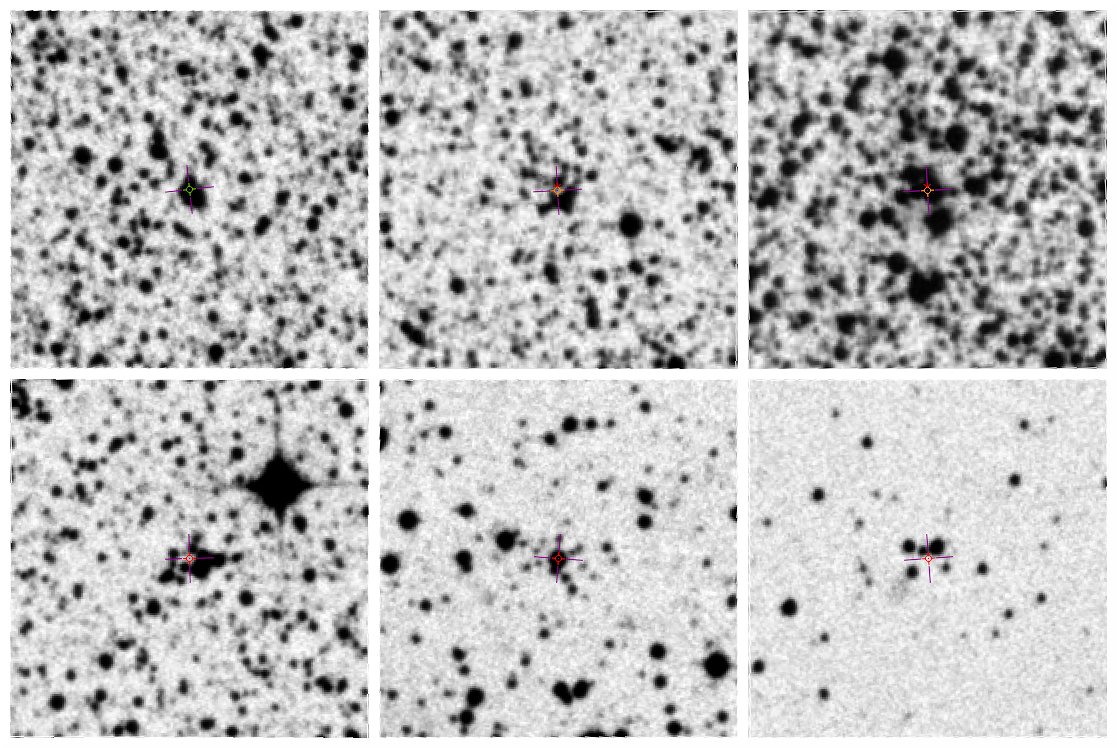}
\caption{Mosaic showing newly identified clusters and candidates,
obtained with the Aladin software, in panels of $2^\prime\times 2^\prime$.
North is up, East is left.
From top left in clockwise direction:
SBica~12, SBica~25, SBica~35, BBica~7, BBica~1 and SBica~40.}\label{mosaic}
\end{figure*}

In Figure~\ref{mosaic} we show examples of newly identified clusters and candidates 
in the SMC main body and the Bridge.
The mosaic shows $2^\prime\times 2^\prime$ images with MAMA $J$ (Blue).
From top left in clockwise direction:
SBica~12 is compact and barely resolved; SBica~25 is more resolved; SBica~35 is compact.
BBica~7 is a small cluster or a cluster core. BBica~1 suggests dissolution, and SBica~40 is small and loose.

\section{Gaia data}

\addtocounter{figure}{-1}
% Fig. 11 / A2
\begin{figure*}
\centering
\includegraphics[width=0.25\linewidth]{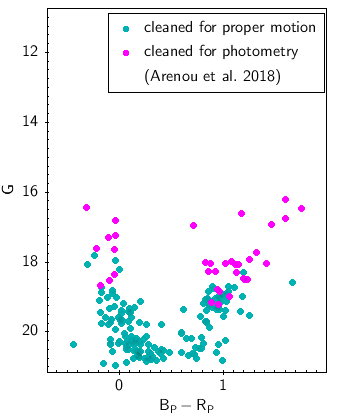}
\includegraphics[width=0.25\linewidth]{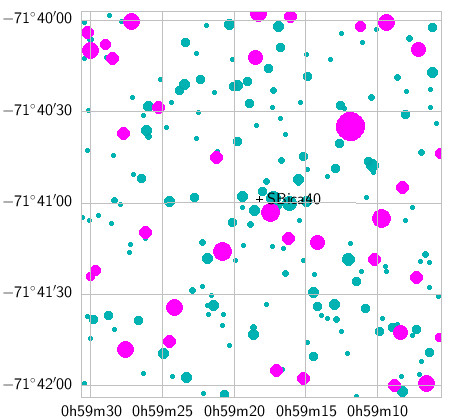}
\includegraphics[width=0.25\linewidth]{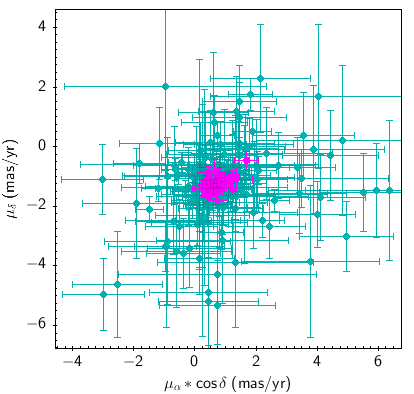}
\caption{CMD (right), sky chart (middle) and VPD (right) of $2^\prime \times 2^\prime$ sample around candidate cluster SBica40, extracted from {\it Gaia} catalog using quality constraints for photometry (magenta) or proper motions (cyan) usage \citep{2018A&A...616A..17A}.}
\label{gaiafig1}
\end{figure*}

Photometry and astrometry from the {\it Gaia} second data release (DR2)
were employed in a attempt to characterise some of the newfound
cluster candidates.
For this purpose, we have used Vizier\footnote{\url{http://vizier.u-strasbg.fr/}}
to extract data inside a $2^\prime\times 2^\prime$ area centred
on prominent SBica~40, matching the angular dimensions of Figure~\ref{mosaic}.
%the finding charts of these targets shown in appendix [include reference here]. 

In order to properly filter bad quality data for photometry purposes,
we have followed the recommendations in \citet{2018A&A...616A..17A},
using their equations 1 and 2 in order to remove poor astrometric solutions,
spurious sources and calibration problems.
These filters have been consistently applied by many authors to produce
reliable photometric analysis.
On the other hand, when only a proper motion analysis is needed,
equations 1 and 3 are recommended instead,
as these filters will retain a much larger fraction of the catalog and still
be useful for astrometric purposes. 

Figure~\ref{gaiafig1} presents the analysis of the Gaia DR2 data for the SBica~40 area.
It shows the resulting cleaned samples of {\it Gaia} data extracted around
cluster candidate SBica~40, aiming at characterising its stellar population.
Although the proper motion sample has a significant number of stars,
it can be seen that its uncertainties on the vector-point diagram (VPD)
are too large to discriminate individual cluster movement from that of
the general LMC and Galactic fields.
Additionally, the distribution of the sample cleaned for photometry in
both the color-magnitude diagram (CMD) and on the sky chart is not sufficient
to carry out a proper analysis of the target.

This analysis was also carried out in more populous clusters of the SMC,
yielding similar results. Therefore, we concluded that the {\it Gaia} data
is not suitable for carrying out a preliminary analysis of such faint clusters.

\section{Possible Extended Magellanic System Clusters and Satellite Dwarf Galaxies}
\label{AppTable4}

% Table 4 / B1
% \clearpage
% \begin{landscape}
% \begin{rotatetable}
% \tabletypesize
% \scriptsize
% \begin{splitdeluxetable*}{lrrcrrrrBccl}
% Como dá para ver, tentei de tudo para manter a tabela inteira, mas nada funcionou
\begin{deluxetable*}{lrrcrrrrccl}
  \rotate
  \tablecaption{Possible Extended Magellanic System Clusters and
    Satellite Dwarf Galaxies\label{tab:landmarks}.}
\tablewidth{0pt}
\tabletypesize{\scriptsize}
% Options fontsize: \small (11pt), \footnotesize (10pt), or \scriptsize (8pt)
\tablehead{
\colhead{Designation(s)} & \colhead{J2000 R. A.} & \colhead{J2000 Dec.} &
\colhead{Class} & \colhead{$D$} & \colhead{$d$} & \colhead{$d_\odot$} &
\colhead{$M_V$} & \colhead{kin.} & \colhead{References$^\dagger$}
& \colhead{Comments} \\
\colhead{} & \colhead{(hh:mm:ss:d)} & \colhead{(Deg:$\arcmin$:$\arcsec$)} & 
\colhead{} & \colhead{($\arcmin$)} & \colhead{($\arcmin$)} & \colhead{(kpc)} &
\colhead{(mag)} & \colhead{} & \colhead{} & \colhead{}
}
\decimalcolnumbers
\startdata
Kim 2, Indus 1, Indus I, DES J2108.8-5109      & 21:08:50.0 & -51:09:49 &    UFC &  2.8 &  2.8 & 100 &  1.3  & n &          (1)          & MW halo, MC origin? \\
DES 3                                          & 21:40:13.2 & -52:32:30 &    UFC &  2.0 &  2.0 &  76 & -1.9  & n &          (4)          & MW halo \\
Grus II, DES J2204-4626                        & 22:04:04.8 & -46:26:24 &     FG & 12.0 & 12.0 &  53 & -3.9  & y &     (1), (12), (13)   & UFC? MC satellite? \\
Tuc II, Tucana II, Tucana 2, DES J2251.2-5836  & 22:51:55.1 & -58:34:08 &     FG & 20.0 & 20.0 &  57 & -3.8  & y &     (1), (10), (13)   & less prob. LMC sat., trailing LMC? \\
Gru I, Grus 1, Grus I                          & 22:56:42.4 & -50:09:48 &     FG &  3.6 &  3.6 & 120 & -3.4  & y &      (1), (8), (10)   & MW Halo, MC origin? Trailing LMC \\
Tuc V, Tucana V, DES J2337-6316                & 23:37:24.0 & -63:16:12 &    UFC &  2.0 &  2.0 &  55 & -1.6  & n &      (1), (6), (13)   & related to the SMC, dissolving? \\
Phe II, Phe 2, Phoenix II, DES J2339.9-5424    & 23:39:58.3 & -54:24:18 &    UFC &  2.2 &  2.2 &  81 & -2.74 & y & (1), (8), (11), (13)  & UFG? former LMC?, LMC sat.? VPO? \\
Tuc III, Tucana III, DES J2356-5935            & 23:56:25.8 & -59:35:00 &    UFG & 12.0 & 12.0 &  25 & -3.4  & y &      (1), (7), (10)   & MC Satellite? \\
Tuc IV, Tucana IV, DES J0002-6051              &  0:02:55.2 & -60:51:00 &    UFG & 18.0 & 18.0 &  48 & -3.5  & y &      (1), (12), (13)  & UFC? MC Satellite: LMC \\
DES 1, DES J0034-4902                          &  0:33:59.8 & -49:07:47 &    UFC &  8.0 &  8.0 &  74 & -1.42 & n &           (6)         & related to the SMC \\
SMCNOD                                         &  0:47:59.9 & -64:48:02 & debris &  360 &  180 &  62 & -7.7  & n &           (9)         & TDG? disrupted SMC satellite \\
Eri III, Eri 3, Eridanus III, DES J0222.7-5217 &  2:22:45.5 & -52:17:05 &    UFC &  2.5 &  2.5 &  91 & -2.07 & y &        (6), (13)      & MC sat., LMC? \\
Hydrus I, Hydrus 1                             &  2:29:33.4 & -79:18:32 &     FG & 13.0 & 13.0 &  28 & -4.7  & y &        (1), (10)      & MW halo, LMC satellite. MC origin? \\
Hor I, Hor 1, Horologium I, DES J0255.4-5406   &  2:55:31.7 & -54:07:08 &     FG &  2.6 &  2.6 &  68 & -3.58 & y & (1), (8), (10), (13)  & LMC satellite \\
Torrealba 1, To 1                              &  3:44:19.8 & -69:25:21 &    UFC &  0.6 &  0.6 &  44 & -1.6  & n &          (4)          & LMC halo? Bridge? Stripped? \\
Hor II, Horologium II                          &  3:16:32.1 & -50:01:05 &    UFG & 19.0 & 19.0 &  78 & -2.1  & y & (1), (5), (11), (13)  & pair w Hor I? LMC satellite \\
Ret II, Reticulum II, Ret 2, DES J0335.6-5403  &  3:35:47.8 & -54:02:48 &    UFG &  7.5 &  7.5 &  30 & -2.7  & y & (1), (10), (12), (13) & less probable LMC satellite \\
Ret III, Reticulum III, DES J0345-6026         &  3:45:26.4 & -60:27:00 &    UFG &  4.8 &  4.8 &  92 & -3.4  & y &    (1), (11), (13)    & UFC? LMC Satellite \\
Pic I, Pictor I, Pictor 1, DES J0443.8-5017    &  4:43:47.4 & -50:16:59 &    UFC &  1.8 &  1.8 & 110 & -2.05 & y &     (1), (8), (13)    & LMC satellite \\
OGLL 863$^\ddagger$, DES 4                      &  5:28:22.8 & -61:43:26 &    UFC &  1.7 &  1.7 &  31 & -1.1  & n &       (14), (4)       & in the LMC, GC? OC? UFG? \\
OGLL 874$^\ddagger$, DES 5                      &  5:10:01.1 & -62:34:49 &    UFC &  0.4 &  0.4 &  25 &  0.3  & n &       (14), (4)       & in the LMC \\
OGLL 845$^\ddagger$, Gaia 3                     &  6:20:14.2 & -73:24:52 &    UFC &  1.1 &  1.1 &  48 & -3.3  & n &       (14), (4)       & in LMC: 1.3Gyr, $\left[{\rm Fe}/{\rm H}\right]=-1.8$ \\
SMASH 1                                        &  6:20:59.9 & -80:23:45 &    UFC &  5.5 &  5.5 &  57 & -1.0  & n &          (3)          & LMC cluster. LMC halo? \\
Pic II, Pictor II, MagLiteS J0644-5953         &  6:44:43.2 & -59:53:49 &    UFG &  7.6 &  7.6 &  45 & -3.2  & n &          (1)          & LMC Satellite, LMC origin \\
Car II, Carina II                              &  7:36:25.6 & -57:59:57 &     FG & 17.0 & 17.0 &  36 & -4.5  & y &        (1), (10)      & LMC satellite \\
Car III, Carina III                            &  7:38:31.2 & -57:53:59 &    UFG &  7.5 &  7.5 &  28 & -3.4  & y &        (1), (10)      & LMC satellite \\
Ant II, Ant 2, Antlia II, Antlia 2             &  9:35:32.8 & -36:46:03 &     FG &  150 &  150 & 130 & -8.5  & y &          (2)          & MW sat., LMC Leading Arm? debris? \\ % or tidal debris
\enddata
\tablecomments{The columns give, respectively: one or more designations;
right ascension and declination in J2000 epoch; major (D) and minor (d) axes
in arcmin; distance to the Sun (d$_{\odot}$) in kpc; absolute magnitude $M_V$;
whether the object has studies about its kinematics or orbits (yes or no);
references list and comments. \\
$^\dagger$The numbers in the references list correspond to:
 (1) \citet{2018ApJ...867...19K}; (2) \citet{2019MNRAS.488.2743T}, (3) \citet{2016ApJ...830L..10M};
 (4) \citet{2019MNRAS.484.2181T}; (5) \citet{2015ApJ...808L..39K}; (6) \citet{2018ApJ...852...68C};
 (7) \citet{2018MNRAS.481.3148E}; (8) \citet{2018arXiv180902259J}; (9) \citet{2017MNRAS.468.1349P};
 (10) \citet{2018AA...619..103F}; (11) \citet{2019AA...623A.129F}; (12) \citet{2018AA...620..155M};
 (13) \citet{2019ApJ...875...77P}; and (14) \citet{2016AcA....66..255S}. \\
$^\ddagger$OGLL are OGLE LMC objects from \citet{2016AcA....66..255S}.\\}
\end{deluxetable*}

Table~\ref{tab:landmarks} lists the faint and ultra-faint clusters and
galaxies (UFC, UFG, FG) populating the Extended Magellanic System (EMS).
These neighbours include satellites, captures, dissolving and co-moving
objects in the vicinity of the Clouds.
Their main characteristics (e.g. position, type, size, distance, brightness)
and references are provided.

\newpage
\thispagestyle{empty}

\end{document}